\documentclass[aps,prd,preprintnumbers,showpacs,showkeys,nofootinbib,
superscriptaddress,fleqn,floatfix,tightenlines,10pt]{revtex4-1}
\usepackage{amsmath,amsfonts,amssymb,amscd,amsxtra,amsthm}
\usepackage{graphicx}  % Include figure files
\usepackage{epstopdf}
\usepackage{dcolumn}  % Align table columns on decimal point 
\usepackage{bm}          % bold math
\usepackage{slashed}
\usepackage{cancel}
\usepackage{float}
\usepackage{mathtools}
\usepackage{amsbsy}
\usepackage{amstext}

\usepackage[utf8]{inputenc} 
\usepackage{booktabs}
\usepackage[normalem]{ulem} % \sout{old text} for strikeout
\usepackage[dvipsnames]{xcolor} % For blue in-text comments and
\usepackage{tabularx}
\usepackage{enumitem}  
\usepackage{array} 
\usepackage{slashed}
\usepackage{tikz}
\usepackage{float}
\usepackage{multirow}
\renewcommand\sout{\bgroup \color{red} \ULdepth=-.5ex \ULset}

\makeatletter

\begin{document}  
\preprint{INHA-NTG-03/2018}
\title{Electromagnetic form factors of singly heavy baryons in the
  self-consistent SU(3) chiral quark-soliton model}
%--------------------------------------------------
%--------------------------------------------------
\author{June-Young Kim}
\email[E-mail: ]{junyoung.kim@inha.edu}
\affiliation{Department of Physics, Inha University, Incheon 22212,
Republic of Korea}

\author{Hyun-Chul Kim}
\email[E-mail: ]{hchkim@inha.ac.kr}
\affiliation{Department of Physics, Inha University, Incheon 22212,
Republic of Korea}
\affiliation{School of Physics, Korea Institute for Advanced Study 
  (KIAS), Seoul 02455, Republic of Korea}
%--------------------------------------------------
\date{\today}
\begin{abstract}
The self-consistent chiral quark-soliton model is a relativistic pion
mean-field approach in the large $N_c$ limit, which describes both
light and heavy baryons on an equal footing. In the limit of the
infinitely heavy mass of the heavy quark, a heavy baryon can be
regarded as $N_c-1$ valence quarks bound by the pion mean fields,
leaving the heavy quark as a color static source. The structure of the
heavy baryon in this scheme is mainly governed by the light-quark
degrees of freedom. Based on this framework, we evaluate the
electromagnetic form factors of the lowest-lying 
heavy baryons. The rotational $1/N_c$ and strange current quark mass
corrections in linear order are considered. We discuss the electric
charge and magnetic densities of heavy baryons in comparison with
those of the nucleons. The results of the electric charge radii of the
positive-charged heavy baryons show explicitly that the heavy baryon
is a compact object. The electric form factors are presented. The form
factor of $\Sigma_c^{++}$ is compared with that from a lattice QCD. We
also discuss the results of the magnetic form factors. The magnetic
moments of the baryon sextet with spin 1/2 and the magnetic radii are
compared with other works and the lattice data.  
\end{abstract}
\pacs{}
\keywords{Heavy baryons, electromagnetic form factors, pion mean
  fields, the chiral quark-soliton model} 
\maketitle
%--------------------------------------------------
\section{Introduction}
A baryon can be viewed as $N_c$ valence quarks bound by the meson
mean field~\cite{Witten:1979kh,Witten:1983tx} in the $1/N_c$ expansion
within quantum chromodynamics (QCD), where $N_c$ denotes the number of
colors. Witten showed explicitly in his seminal
paper~\cite{Witten:1979kh} that in two-dimensional QCD the baryon 
can be considered as a bound state of $N_c$ valence quarks by 
the meson mean fields in the Hartree approximation. Since the mass of
the nucleon is proportional to $N_c$ and the meson-loop fluctuations 
are suppressed by $1/N_c$, such a mean-field approach is valid in the
large $N_c$ limit. The chiral quark-soliton model
($\chi$QSM)~\cite{Diakonov:1987ty, Wakamatsu:1990ud, Christov:1995vm,
  Diakonov:1997sj} was developed based on this idea. In the large
$N_c$ limit, the presence of the $N_c$ valence quarks produces the
chiral mean fields coming from the polarization of a Dirac sea that in
turn influences self-consistently the valence quarks. In this picture,
the baryon arises as a soliton that consists of the $N_c$ valence
quarks. A very important feature of the chiral mean field or the
soliton is hedgehog symmetry. The $\chi$QSM described successfully
various properties of the baryon octet and decuplet such as the
electromagnetic (EM) properties~\cite{Kim:1995mr, Kim:1995ha,
  Wakamatsu:1996xm, Kim:1997ip, Silva:2013laa}, axial-vector form 
factors~\cite{Silva:2005fa}, tensor charges and form
factors~\cite{Kim:1995bq, Kim:1996vk, Pobylitsa:1996rs,
  Schweitzer:2001sr, Ledwig:2010tu, Ledwig:2010zq}, 
semileptonic decays~\cite{Kim:1997ts, Ledwig:2008ku, Yang:2015era},
parton distributions~\cite{Diakonov:1996sr, Diakonov:1997vc, 
  Wakamatsu:1997en}, and so on.     

Very recently, it was shown that singly heavy baryons can be
considered as $N_c-1$ valence quarks bound by the pion mean
field~\cite{Yang:2016qdz}, being motivated by
Diakonov~\cite{Diakonov:2010tf}. In the limit of the infinitely heavy
quark mass ($m_Q\to \infty$), the spin of the heavy quark $\bm{J}_Q$
is conserved, which leads to the conservation of the spin of the
light-quark degrees of freedom: $\bm{J}=\bm{J}'-\bm{J}_Q
$~\cite{Isgur:1989vq, Georgi:1990um}. In
this limit, the heavy quark inside a singly heavy baryon can be
merely regarded as a static color source, which means that the heavy
quark is required only to make the heavy baryon a color singlet. Thus,     
the flavor $\mathrm{SU_{f}(3)}$ representations of the 
lowest-lying heavy baryons are given by $\bm{3}\otimes \bm{3}=
\overline{\bm{3}}\oplus \bm{6}$, of which the baryon antitriplet has
$J=0$ and $J'=1/2$, whereas the sextet has $J=1$. The light quarks with
$J=1$ being coupled to the spin of the heavy quark $J_Q=1/2$, the
baryon sextet have spins $1/2$ and $3/2$. So, there are two
representations with spin $1/2$ and $3/2$ in the baryon sextet, which
are degenerate in the limit of $m_Q\to \infty$. The hyperfine
spin-spin interaction will lift the degeneracy between these states
with different spins~\cite{Yang:2016qdz}.  

In the $\chi$QSM, we can apply basically the same formalism on the
singly heavy baryons, which have been developed for the description of 
light baryons. Considering the heavy quark as a static color source,
the heavy baryon can be regarded as a system of the $N_c-1$ valence
quarks with the heavy quark stripped off from the valence level. In
the case of the light baryons, the collective Hamiltonian is
constrained by the right hypercharge $Y'=N_c/3$ imposed by the $N_c$
valence quarks, which selects the lowest allowed 
$\mathrm{SU_{f}(3)}$ representations such as the octet 
($\bm{8}$) and the decuplet ($\bm{10}$). However, when it comes to the
singly heavy baryons, this constraint should be changed to be
$Y'=(N_c-1)/3$ because of the $N_c-1$ valence quarks inside a heavy
baryon and yields the antitriplet ($\overline{\bm{3}}$) and the
sextet ($\bm{6}$) as the lowest allowed representations. In addition,   
we need to modify the valence parts of all moments of inertia and
quark densities for the calculation of any form factors. This
extension of the $\chi$QSM was rather successful in describing  
the masses of the lowest-lying singly heavy baryons in both the
charmed and bottom sectors, and the mass of the $\Omega_b^{*}$ 
was predicted~\cite{Yang:2016qdz, Kim:2018xlc}. Moreover, newly found 
$\Omega_c$ resonances~\cite{Aaij:2017nav} were well explained and
classified. In particular, two of the $\Omega_c$ resonances were
interpreted as exotic baryons belonging to the antidecapentaplet 
($\overline{\bm{15}}$) with their narrow widths correctly
reproduced~\cite{Kim:2017jpx, Kim:2017khv}.   

In the present work, we want to investigate the EM
properties of the lowest-lying singly heavy baryons with spin
$1/2$. Assuming that the mass of the heavy quark is infinitely heavy, 
we will show that the main features of the EM form factors are
governed by the $N_c-1$ light quarks. Of course, the electric form
factor requires a certain contribution from the heavy quark such that
the charge of the heavy baryons should be correctly reproduced. Since the EM
current is decomposed into the light and heavy parts, the heavy-quark 
contribution can be treated separately. Its effects on the electric form
factors are just the constant ones within the present framework, which
indicates that the heavy quark is considered to be a structureless
particle. This is a natural consequence, because we deal with the
heavy quark just as a static color source. 
Since any form factor in QCD should decrease rapidly as the
square of the momentum transfer increases because of gluon exchanges
between the quarks that constitutes a baryon, the present 
picture of the electric form factors may be put into
question. However, keeping in mind that the $\chi$QSM is the
low-energy effective theory of the nucleon, we still can treat the
heavy quark as a static one in the limit of $m_Q\to \infty$, as far as
the momentum transfer remains much smaller than the heavy-quark mass,
i.e., $q^2\ll m_Q^2$.  
We will show that this approach indeed produces reasonable results for
the electric form factors of the heavy baryons in comparison with the
lattice data~\cite{Can:2013tna}. In the case of the magnetic form
factors, the situation is even better. In the limit of $m_Q\to
\infty$, the effects of the heavy quark vanishes, since the magnetic
moment of the heavy quark is proportional to the inverse of the
heavy-quark mass. Thus, the magnetic form factors of the singly heavy
baryons are solely governed by the light quarks.    

We sketch the structure of the present paper as follows: In Sec.
II, we show how to compute the EM form factors of the
heavy baryons within the $\chi$QSM. In Sec. III, we present and
discuss the numerical results of the EM form factors and the
corresponding charge and magnetic radii. We also examine the effects
of the $\mathrm{SU_{f}(3)}$ symmetry breaking.
The final section is devoted to the summary and conclusions. 

%=================
\section{General formalism}
%=================
The EM current including the heavy quark (the charm quark
or the bottom quark) is expressed as 
\begin{align}
  \label{eq:LHcurrent}
J_\mu (x) = \bar{\psi} (x) \gamma_\mu \hat{\mathcal{Q}} \psi(x) + e_{Q}
  \bar{\Psi}   \gamma_\mu \Psi,  
\end{align}
where $\hat{\mathcal{Q}}$ denotes the charge operator in 
$\mathrm{SU_{f}(3)}$, defined by
\begin{align}
 \label{eq:chargeOp}
\hat{\mathcal{Q}} =
  \begin{pmatrix}
   \frac23 & 0 & 0 \\ 0 & -\frac13 & 0 \\ 0 & 0 & -\frac13
  \end{pmatrix} = \frac12\left(\lambda_3 + \frac1{\sqrt{3}}
                                                  \lambda_8\right). 
\end{align}
Here, $\lambda_3$ and $\lambda_8$ are the flavor SU(3) Gell-Mann
matrices. The $e_Q$ in the second part of the EM current
in Eq.~(\ref{eq:LHcurrent}) stands for the heavy-quark charge, which
is given as $e_c=2/3$ for the charm quark or as $e_b=-1/3$ for the
bottom quark.  Since the magnetic form factor of a heavy quark is
proportional to the inverse of the corresponding heavy-quark mass,
i.e., $\bm{\mu} \sim (e_Q/m_Q) \bm{\sigma}$, we can ignore the
contribution from the heavy quark current in the limit of $m_Q\to
\infty$.  However, we need to keep the second term in
Eq.~\eqref{eq:LHcurrent} when we compute the electric form factors. 
The EM form factors of the spin-$1/2$ baryons are defined
by the matrix element of the EM current 
\begin{align}
\langle B,\,p' | J_\mu(0) |B, \,p\rangle = \overline{u}_{B}(p',\,\lambda') 
  \left[\gamma_\mu F_1(q^2) + i\sigma_{\mu\nu} \frac{q^\nu}{2M_N}
  F_2(q^2)\right] u_B(p,\,\lambda),   
\label{eq:MatrixEl1}
\end{align}
where $M_N$ is the mass of the nucleon. The $q^2$ stands for the
four-momentum transfer $q^2=-Q^2$ with $Q^2 >0$. $u_B(p,\,\lambda)$
denotes the Dirac spinor with the momentum $p$ and the helicity
$\lambda$. The Dirac and Pauli form factors $F_1(Q^2)$ and $F_2(Q^2)$
can be written in terms of the 
Sachs EM form factors, $G_E(Q^2)$ and $G_M(Q^2)$ 
\begin{align}
G_E^B(Q^2) &= F_1^B (Q^2) - \tau F_2^B (Q^2), \cr
G_M^B(Q^2) &= F_1^B (Q^2) + F_2^B(Q^2),
\end{align}
with $\tau=Q^2/4M_N^2$. In the Breit frame, the Sachs form factors are 
related to the time and space components of the EM current,
respectively,
\begin{align}
 G_E^B(Q^2) &= \int \frac{d\Omega_q}{4\pi} \langle B,\,p' |J_0(0)
              |B,\,p\rangle,\cr 
G_M^B(Q^2) &=  3 M_N \int \frac{d\Omega_q}{4\pi}
             \frac{q^i\epsilon_{ik3}}{i|\bm{q}|^2} \langle B,\,p'
             |J^k(0) |B,\,p\rangle.  
\end{align}
Thus, once we compute the matrix elements of the EM current, we can
directly derive the EM form factors. Note that we consider the
heavy-quark part separately. 

The SU(3) $\chi$QSM is characterized by the following low-energy
effective partition function in Euclidean space
\begin{align}
\label{eq:partftn}
\mathcal{Z}_{\chi\mathrm{QSM}} = \int \mathcal{D}\psi \mathcal{D}
  \psi^\dagger \mathcal{D} U \exp\left[-\int d^4 x \psi^\dagger i D(U)
  \psi\right]  = \int \mathcal{D} U \exp (-S_{\mathrm{eff}}),  
\end{align}
where $\psi$ and $U$ represent the quark and pseudo-Nambu-Goldstone
boson fields, respectively. 
The $S_{\mathrm{eff}}$ is the effective chiral action,  
\begin{align}
S_{\mathrm{eff}}(U) \;=\; -N_{c}\mathrm{Tr}\ln iD(U)\,,
\label{eq:echl}
\end{align}
where $\mathrm{Tr}$ stands for the generic trace operator running over 
spacetime and all relevant internal spaces. The $N_c$ is the number
of colors, and $D(U)$ the Dirac differential operator is defined by 
\begin{align}
D(U) \;=\; 
\gamma_{4}(i\rlap{/}{\partial} - \hat{m} - MU^{\gamma_{5}}) =
-i\partial_{4} + h(U) - \delta m,
\label{eq:Dirac}  
\end{align}
where $\partial_4$ denotes the Euclidean time derivative.
We assume isospin symmetry,
i.e., $m_{\mathrm{u}}=m_{\mathrm{d}}$. We define the average mass of
the up and down quarks by $\overline{m}=(m_{\mathrm{u}} +
m_{\mathrm{d}})/2$. Then, the matrix of the current quark masses is
written as $\hat{m} = \mathrm{diag}(\overline{m},\, \overline{m},\,
m_{\mathrm{s}}) = \overline{m} +\delta m$. $\delta m$ is written as 
\begin{align}
\delta m  \;=\; \frac{-\overline{m} + m_{s}}{3}\gamma_{4}\bm{1} +
\frac{\overline{m} - m_{s}}{\sqrt{3}} \gamma_{4} \lambda^{8} =
M_{1}\gamma_{4} \bm{1} + M_{8} \gamma_{4} \lambda^{8}\,,
\label{eq:deltam}
\end{align}
where $M_1$ and $M_8$ are the singlet and octet components of the
current quark masses, expressed, respectively, as 
$M_1=(-\overline{m} +m_{\mathrm{s}})/3 $ and $M_8=(\overline{m}
-m_{\mathrm{s}})/\sqrt{3}$. 
The SU(3) single-quark Hamiltonian $h(U)$ is defined as 
\begin{align}
h(U) \;=\;
i\gamma_{4}\gamma_{i}\partial_{i}-\gamma_{4}MU^{\gamma_{5}} -
\gamma_{4} \overline{m}\, ,
\label{eq:diracham}  
\end{align}
where $U^{\gamma_5}$ represents the SU(3) chiral field. Since the
hedgehog symmetry constrains the form of the classical pion field as 
$\bm{\pi}(\bm{x}) = \hat{\bm{n}}\cdot \bm{\tau} P(r)$, where $P(r)$ is
the profile function of the soliton, the SU(2) chiral field is written
as   
\begin{align}
U_{\mathrm{SU(2)}}^{\gamma_5} \;=\; \exp(i\gamma^{5}\hat{\bm{n}}\cdot
\bm{\tau} P(r))
\;=\; \frac{1+\gamma^{5}}{2}U_{\mathrm{SU(2)}} +
  \frac{1-\gamma^{5}}{2}U_{\mathrm{SU(2)}}^{\dagger}
\label{eq:embed}
\end{align}
with $U_{\mathrm{SU(2)}}=\exp(i\hat{\bm{n}}\cdot \bm{\tau} P(r))$. 
We now embed the SU(2) soliton into SU(3) by Witten's
ansatz~\cite{Witten:1983tx}  
\begin{align}
U^{\gamma_{5}}(x) \;=\; \left(\begin{array}{lr}
U_{\mathrm{SU(2)}}^{\gamma_{5}}(x) & 0\\
0 & 1
\end{array}\right).
\end{align}
Since we consider the mean-field approximation, we can carry out the
integration over $U$ in Eq.~\eqref{eq:partftn} around the saddle
point ($\delta S_{\mathrm{eff}}/\delta P(r) =0$). This saddle-point
approximation yields the equation of motion that can be solved
self-consistently. The solution provides the self-consistent profile
function $P_c(r)$. 

The matrix elements of the EM current~\eqref{eq:MatrixEl1} can be
computed within the $\chi$QSM by representing them in terms of the
functional integral in Euclidean space,
\begin{align}
\langle B,\,p'| J_\mu(0) |B,\,p\rangle  &= \frac1{\mathcal{Z}}
  \lim_{T\to\infty} \exp\left(i p_4\frac{T}{2} - i p_4'
  \frac{T}{2}\right) \int d^3x d^3y \exp(-i \bm{p}'\cdot \bm{y} + i
  \bm{p}\cdot \bm{x}) \cr
& \hspace{-1cm} \times \int \mathcal{D}U\int \mathcal{D} \psi \int
  \mathcal{D} \psi^\dagger J_{B}(\bm{y},\,T/2) \psi^\dagger(0)
  \gamma_4\gamma_\mu \hat{Q} \psi(0) J_B^\dagger (\bm{x},\,-T/2)
  \exp\left[-\int d^4 z   \psi^\dagger iD(U) \psi\right],  
\label{eq:correlftn}
\end{align}
where the baryon states $|B,\,p\rangle$ and $\langle B,\,p'|$ are, 
respectively, defined by 
\begin{align}
|B,\,p\rangle &= \lim_{x_4\to-\infty}   \exp(i p_4 x_4)
                \frac1{\sqrt{\mathcal{Z}}} \int d^3 x
                \exp(i\bm{p}\cdot \bm{x}) J_B^\dagger
                (\bm{x},\,x_4)|0\rangle,\cr
\langle B,\,p'| &= \lim_{y_4\to\infty}   \exp(-i p_4' y_4)
                \frac1{\sqrt{\mathcal{Z}}} \int d^3 y
                \exp(-i\bm{p}'\cdot \bm{y}) \langle 0| J_B^\dagger 
                (\bm{y},\,y_4).
\end{align}
The heavy baryon current $J_B$ can be constructed from the $N_c-1$
valence quarks
\begin{align}
J_B(x) = \frac1{(N_c-1)!} \epsilon_{i_1\cdots i_{N_c-1}} \Gamma_{JJ_3
  TT_3 Y}^{\alpha_1\cdots \alpha_{N_c-1}} \psi_{\alpha_1 i_1} (x)
  \cdots \psi_{\alpha_{N_c-1} i_{N_c-1}}(x),  
\end{align}
where $\alpha_1\cdots \alpha_{N_c-1}$ represent spin-flavor indices
and $i_1\cdots i_{N_c-1}$ color indices. The matrices $\Gamma_{JJ_3
  TT_3 Y}^{\alpha_1\cdots \alpha_{N_c-1}}$ are taken to consider the
quantum numbers $JJ_3TT_3Y$ of the $N_c-1$ soliton. The creation
operator $J_B^\dagger$ can be constructed in a similar
way. As for the detailed formalism of the zero-mode quantization and
the techniques of computing the baryonic correlation function given in
Eq.~\eqref{eq:correlftn}, we refer to Refs.~\cite{Christov:1995vm,
  Kim:1995mr}. 

Having quantized the soliton, we obtain the collective Hamiltonian as    
\begin{align}
H_{\mathrm{coll}} = H_{\mathrm{sym}} + H_{\mathrm{sb}},   
\end{align}
where
\begin{align}
  \label{eq:Hamiltonian}
H_{\mathrm{sym}} &= M_{\mathrm{cl}} + \frac1{2I_1} \sum_{i=1}^3
                   J_i^2 + \frac1{2I_2} \sum_{p=4}^7 J_p^2,\cr
H_{\mathrm{sb}} &= \alpha D_{88}^{(8)} + \beta \hat{Y} +
  \frac{\gamma}{\sqrt{3}} \sum_{i=1}^3 D_{8i}^{(8)} \hat{J}_i.
\end{align}
$I_1$ and $I_2$ are the soliton moments of inertia. The parameters
$\alpha$, $\beta$, and $\gamma$ for heavy baryons are defined by 
\begin{align}
\alpha=\left (-\frac{\overline{\Sigma}_{\pi N}}{3m_0}+\frac{
  K_{2}}{I_{2}}\overline{Y}  
\right )m_{\mathrm{s}},
 \;\;\;  \beta=-\frac{ K_{2}}{I_{2}}m_{\mathrm{s}}, 
\;\;\;  \gamma=2\left ( \frac{K_{1}}{I_{1}}-\frac{K_{2}}{I_{2}} 
 \right ) m_{\mathrm{s}},
\label{eq:alphaetc}  
\end{align}
where that the three parameters $\alpha$, $\beta$, and $\gamma$ are 
expressed in terms of the moments of inertia $I_{1,\,2}$ and
$K_{1,\,2}$. The valence parts of them are different from those in
the light baryon sector by the color factor $N_c-1$ in place of
$N_c$. The expression of $\overline{\Sigma}_{\pi N}$ is similar to the
$\pi N$ sigma term again except for the $N_c$ factor:
$\overline{\Sigma}_{\pi N} = (N_c-1)N_c^{-1} \Sigma_{\pi N}$. The
detailed expressions for the moments of inertia and
$\overline{\Sigma}_{\pi N}$ are found in Ref.~\cite{Kim:2018xlc}. 

Because of the symmetry-breaking part of the collective Hamiltonian
$H_{\mathrm{sb}}$, the baryon wave functions are no more pure
states but are mixed ones with those in higher SU(3)
representations. Thus the wave functions for the baryon antitriplet
($J=0$) and the sextet ($J=1$) are derived, respectively,
as~\cite{Kim:2018xlc}    
\begin{align}
&|B_{\overline{\bm3}_{0}}\rangle = |\overline{\bm3}_{0},B\rangle + 
p^{B}_{\overline{15}}|\overline{\bm{15}}_{0},B\rangle, \cr
&|B_{\bm6_{1}}\rangle = |{\bm6}_{1},B\rangle +
  q^{B}_{\overline{15}}|{\overline{\bm{15}}}_{1},B 
\rangle + q^{B}_{\overline{24}}|{
{\overline{\bm{24}}}_{1}},B\rangle,
\label{eq:mixedWF1}
\end{align}
with the mixing coefficients
\begin{eqnarray}
p_{\overline{15}}^{B}
\;\;=\;\;
p_{\overline{15}}\left[\begin{array}{c}
-\sqrt{15}/10\\
-3\sqrt{5}/20
\end{array}\right], 
& 
q_{\overline{15}}^{B}
\;\;=\;\;
q_{\overline{15}}\left[\begin{array}{c}
\sqrt{5}/5\\
\sqrt{30}/20\\
0
\end{array}\right], 
& 
q_{\overline{24}}^{B}
\;\;=\;\;
q_{\overline{24}}\left[\begin{array}{c}
-\sqrt{10}/10\\
-\sqrt{15}/10\\
-\sqrt{15}/10
\end{array}\right],
\label{eq:pqmix}
\end{eqnarray}
respectively, in the basis $\left[\Lambda_{Q},\;\Xi_{Q}\right]$ for the
antitriplet and $\left[\Sigma_{Q},\;
  \Xi_{Q}^{\prime},\;\Omega_{Q}\right]$ for the sextets. The 
parameters $p_{\overline{15}}$, $q_{\overline{15}}$, and
$q_{\overline{24}}$ are given by 
\begin{eqnarray}
p_{\overline{15}}
\;\;=\;\;
\frac{3}{4\sqrt{3}}\alpha {I}_{2},
& 
q_{\overline{15}}
\;\;=\;\;
{\displaystyle -\frac{1}{\sqrt{2}}}
\left(\alpha+\frac{2}{3}\gamma\right)
{I}_{2},
& 
q_{\overline{24}}\;\;=\;\;
\frac{4}{5\sqrt{10}}
\left(\alpha-\frac{1}{3}\gamma\right) I_{2}.
\label{eq:pqmix2}
\end{eqnarray}

Having obtained the mixing parameters, we are able to express
explicitly the wave function of a state with flavor $F=(Y,T,T_3)$ and
spin $S=(Y'=-2/3,J,J_3)$ in the representation $\nu$ in terms
of a tensor with two indices, i.e., $\psi_{(\nu;\,
  F),(\overline{\nu};\,\overline{S})}$, one running over the states
$F$ in the representation $\nu$ and the other one over the states
$\overline{S}$ in the representation $\overline{\nu}$. Here, 
$\overline{\nu}$ represents the complex conjugate of the
$\nu$, and the complex conjugate of $S$ is given as
$\overline{S}=(2/3,\,J,\,J_3)$. Since a singly heavy baryon consists
of $N_c-1$ light valence quarks, the constraint imposed on the right
hypercharge should be modified from $\overline{Y}=-Y'=N_c/3$ to
$\overline{Y}=(N_c-1)/3$. Thus, the collective wave function for the
soliton with $(N_c-1)$ valence quarks is written as 
\begin{align}
  \label{eq:SolitonWF1}
\psi_{(\nu;\, F),(\overline{\nu};\,\overline{S})}(R) =
  \sqrt{\mathrm{dim}(\nu)} (-1)^{Q_S} [D_{F\,S}^{(\nu)}(R)]^*,
\end{align}
where $\mathrm{dim}(\nu)$ represents the dimension of the
representation $\nu$ and $Q_S$ a charge corresponding to the baryon
state $S$, i.e., $Q_S=J_3+Y'/2$.  

The complete wave function for a heavy baryon can be constructed by 
coupling the soliton wave function to the heavy quark
\begin{align}
\Psi_{B_{Q}}^{(\mathcal{R})}(R)
 =  \sum_{J_3,\,J_{Q3}} 
C_{\,J,J_3\, J_{Q}\,J_{Q3}}^{J'\,J_{3}'}
\;\mathbf{\chi}_{J_{Q3}}
\;\psi_{(\nu;\,Y,\,T,\,T_{3})(\overline{\nu};\,Y^{\prime},\,J,\,J_3)}(R)
\label{eq:HeavyWF}
\end{align}
where $\chi_{J_{Q3}}$ denote the Pauli spinors and $C_{\,J,J_3\,
  J_{Q}\,J_{Q3}}^{J'\,J_{3}'}$ the Clebsch-Gordan
coefficients. 

The final expression for the electric form factor of a heavy baryon
$B$ can be written as  
\begin{align}
{{G}}^{B}_{E}  (q^{2})= \int d^{3} z j_{0}(|\bm{q}| |\bm{z}|)
  {\mathcal{G}}^{B}_{E} (\bm{z}) + G_E^Q(q^2), 
\label{eq:app1}
\end{align}
where the first part of Eq.~\eqref{eq:app1} is the light-quark
contribution to the electric form factor whereas the second part
corresponds to the pointlike heavy quark.
The electric charge density of a light-quark part can be expressed as    
\begin{align}
{\cal{G}}^{B}_{E} (\bm{z})=& \frac{1}{\sqrt{3}} \langle
  D^{(8)}_{Q8}\rangle_B \mathcal{B}(\bm{z}) -
  \frac{2}{I_{1}} \langle
  D^{(8)}_{Qi} \hat{J}_{i} \rangle_B
  {\cal{I}}_{1}(\bm{z}) -
  \frac{2}{I_{2}} \langle
  D^{(8)}_{Qp} \hat{J}_{p} \rangle_B
  {\cal{I}}_{2}(\bm{z}) \cr 
    & -\frac{4 M_{8}}{I_{1}} \langle D^{(8)}_{8i} D^{(8)}_{Qi}
      \rangle_B (I_{1}{\cal{K}}_{1}(z) - K_{1}{\cal{I}}_{1}(z)) \cr 
 & -\frac{4 M_{8}}{I_{2}} \langle D^{(8)}_{8p} D^{(8)}_{Qp}
   \rangle_B (I_{2}{\cal{K}}_{2}(z) - K_{2}{\cal{I}}_{2}(z)) \cr 
 &-2\left (  \frac{M_1}{\sqrt{3}}\langle D^{(8)}_{Q8} \rangle_B +
   \frac{M_8}{3}\langle D^{(8)}_{8 8}D^{(8)}_{Q8} \rangle_B  \right)
   {\cal{C}}(\bm{z}), 
\label{eq:elecfinal}
\end{align}
where the explicit expressions for the electric densities can be found
in Refs.~\cite{Kim:1995mr,Ledwig:1900ri} with the prefactors of the
valence parts replaced by $N_c-1$. The indices $i$ and $p$ are dummy
ones running over $i=1,\cdots, 3$ and $p=4,\cdots 7$, respectively. In
the present mean-field approach, the heavy-quark contribution to the 
electric form factor is just the constant charge of the corresponding
heavy quark ($e_c=2/3$ or $e_b=-1/3$), because the heavy quark is
assumed to be a static color source and a pointlike particle. Of
course, this is a rather crude approximation but it is still a
reasonable one as far as we consider the electric form factors in low
$Q^2$ regions. Thus, we set $G_E^Q(Q^2)=e_Q$ in the present
work. 

Since the integrations of the densities in Eq.~\eqref{eq:elecfinal}
are given as 
\begin{align}
\int d^3 z \,\mathcal{B}(\bm{z}) ={N_{c}},\;\;\;
\frac1{I_i}\int d^3 z \,\mathcal{I}_i (\bm{z}) = 1,\;\;\;  
\frac1{K_i}\int d^3 z \,\mathcal{K}_i (\bm{z}) = 1,\;\;\;  
\int d^3 z \,\mathcal{C}(\bm{z}) = 0,\;\;\;  
\end{align}
and $G_E^Q(0) = e_Q$, the electric form factor $G_E^B$ at $Q^2=0$
turns out to be the charge of the corresponding heavy baryon. 

The expression for the magnetic moment form factor of a baryon $B$ is
written as  
\begin{align}
{{G}}^{B}_{M}  (q^{2})=\frac{ M_{N}}{|\bm{q}|} \int d^{3} z
  \frac{j_{1}(|\bm{q}| |\bm{z}|)}{ |\bm{z}|} {\cal{G}}^{B}_{M}
  (\bm{z}) , 
\label{eq:magfinal}
\end{align}
where the corresponding density of the magnetic form factors is given
by 
\begin{align}
{\cal{G}}^{B}_{M}(\bm{z}) &=  \langle D^{(8)}_{Q3 }
  \rangle_B  \left(  {\cal{Q}}_{0} (\bm{z})  + \frac{1}{I_{1}}
 {\cal{Q}}_{1} (\bm{z}) \right) -  \frac{1}{\sqrt{3}} \langle
 D^{(8)}_{Q 8}J_{3} \rangle_B \frac{1}{I_{1}}
 {\cal{X}}_{1} (\bm{z}) - \langle
 d_{pq3} D^{(8)}_{Qp} J_{q} \rangle_B
 \frac{1}{I_{2}}  {\cal{X}}_{2} (\bm{z})   \cr 
& + \frac{2}{\sqrt{3}} M_{8} \langle D^{(8)}_{83}
  D^{(8)}_{Q8} \rangle_B \left(\frac{K_{1}}{I_{1}}{\cal{X}}_{1}
  (\bm{z}) -   {\cal{M}}_{1} (\bm{z})\right)  
  +2 M_{8} \langle  d_{pq3}  D^{(8)}_{8p}
  D^{(8)}_{Qq} \rangle_B \left(\frac{K_{2}}{I_{2}}{\cal{X}}_{2} (\bm{z})
     -    {\cal{M}}_{2} (\bm{z})\right)  \cr 
& - 2  \left( M_{1} \langle D^{(8)}_{Q3} \rangle_B +
  \frac{1}{\sqrt{3}} M_{8} \langle D^{(8)}_{88} D^{(8)}_{Q3}
  \rangle_B  \right) {\cal{M}}_{0} (\bm{z}). 
\label{eq:magden}
\end{align}
The indices $p$ and $q$ are the dummy indices running over $4\cdots
7$. The explicit forms for the magnetic densities can be found in
Ref.~\cite{Kim:1995mr, Ledwig:1900ri} with the prefactors of the
valence parts replaced by $N_c-1$. The matrix elements of the
collective operators are explicitly given in Appendix~\ref{app:B}. The
magnetic form factor at $Q^2=0$ produces the magnetic moment of the
corresponding baryon. So, it is convenient to express a collective
operator for the magnetic moments~\cite{Yang:2018uoj} as 
\begin{align}
\hat{\mu} & =  
\;\;w_{1}D_{\mathcal{Q}3}^{(8)}
\;+\;w_{2}d_{pq3}D_{\mathcal{Q}p}^{(8)}\cdot\hat{J}_{q}
\;+\;\frac{w_{3}}{\sqrt{3}}D_{\mathcal{Q}8}^{(8)}\hat{J}_{3} \cr
& +\frac{w_{4}}{\sqrt{3}}d_{pq3}D_{\mathcal{Q}p}^{(8)}D_{8q}^{(8)}
+w_{5}\left(D_{\mathcal{Q}3}^{(8)}D_{88}^{(8)}+D_{\mathcal{Q}8}^{(8)}
                  D_{83}^{(8)}\right) 
\;+\;w_{6}\left(D_{\mathcal{Q}3}^{(8)}D_{88}^{(8)}-
                  D_{\mathcal{Q}8}^{(8)}D_{83}^{(8)}\right),   
\label{eq:magop}
\end{align}
where the dynamical coefficients $w_i$ can be found in
Appendix~\ref{app:magmom}. The results of the magnetic moments and
$w_i$ are compared with those from the model-independent
analysis~\cite{Yang:2018uoj} also in Appendix~\ref{app:magmom}.
\section{Results and discussion}
We are now in a position to discuss the results from the present
work. We first briefly mention how to fix the parameters of the
model. We refer to Refs.~\cite{Christov:1995vm, Kim:1995mr} for a 
detailed explanation of numerical methods. The only free parameter of
the $\chi$QSM is the dynamical quark mass $M$ of which the numerical
value was already fixed by computing various form factors of the
nucleon. Its most preferable value is $M=420$ MeV. Nevertheless, we have
checked whether the present results are sensitive to it with $M$
varied from 400 to 450 MeV. All the form factors presented in this
work are rather insensitive to the value of $M$, so we choose the
value for the best fit, i.e., $M=420$ MeV as in 
the light-baryon sector case~\cite{Christov:1995vm, Kim:1995mr,
  Silva:2005fa,Ledwig:2010tu, Ledwig:2010zq, Ledwig:2008ku}. Note that
the same value of $M$ was selected also for the mass splitting of the
heavy baryons~\cite{Kim:2018xlc}. There are yet another parameters in
the $\chi$QSM: the average mass of the current up and down quarks
$\overline{m}$, the strange current quark mass $m_{\mathrm{s}}$, and
the cutoff parameter $\Lambda$ of the proper-time regularization. The
value of $\overline{m}$ was fixed to be $\overline{m}=6.131$ MeV by
reproducing the pion mass whereas the cutoff parameter is determined
by reproducing the pion decay constant $f_\pi=93$ MeV. 

The strange current quark mass can be in principle taken from its
canonical value $m_{\mathrm{s}}=150$ MeV which was obtained by
reproducing the kaon mass in the model. However, $m_{\mathrm{s}}=180$
MeV was used for the calculation of the form factors and other
properties of the $\mathrm{SU_{\mathrm{f}}(3)}$ light baryons
effectively. Very recently, the dependence of the mass splittings of
heavy baryons on $m_{\mathrm{s}}$ were examined within the same
framework of the $\chi$QSM~\cite{Kim:2018xlc} and the best values of
$m_{\mathrm{s}}$ were obtained to be $m_{\mathrm{s}}=174$ MeV and
$m_{\mathrm{s}}=166$ MeV for the mass splittings of the charmed
baryons and the bottom baryons, respectively. Thus, instead of using
the previous value 180 MeV, we will use the same values of
$m_{\mathrm{s}}$ as obtained in Ref.~\cite{Kim:2018xlc} for
consistency, regarding $m_{\mathrm{s}}$ as an effective mass. However,
the EM form factors of the heavy baryons show rather tiny dependence
on the numerical value of $m_{\mathrm{s}}$, so the difference of the
$m_{\mathrm{s}}$ value does not affect the results at all.   

\subsection{Electric form factors of the baryon antitriplet and
  sextet with spin 1/2}
The electric form factor of a baryon at $Q^2=0$ is the same as its
corresponding charge. Integrating the electric charge density
of the baryon given in Eq.~\eqref{eq:elecfinal} over three-dimensional
space, one obtains the corresponding charge. In fact, the collective
charge operator is found from Eq.~\eqref{eq:elecfinal}:  
\begin{align}
\hat{Q} = \frac{N_c}{2\sqrt{3}} D_{38}^{(8)} + \frac{N_c}{6}
  D_{88}^{(8)} + 
  \sum_{i=1}^7 D_{3i}^{(8)} \hat{T}_i + \frac1{\sqrt{3}}
  \sum_{i=1}^7 D_{8i}^{(8)} \hat{T}_i,    
\end{align}
where $\hat{T}_i$ are the generators of $\mathrm{SU(3)}$ group. Using
the relations 
\begin{align}
\hat{T}_8=\frac{N_c}{2\sqrt{3}},\;\;\; \hat{T}_3=\sum_{i=1}^8
  D_{3i}^{(8)} \hat{T}_i,\;\;\;    \hat{Y} = \frac{2}{\sqrt{3}}
  \sum_{i=1}^8 D_{8i}^{(8)} \hat{T}_i,   
\end{align}
Then we find the well-known Gell-Mann-Nishijima formula in
$\mathrm{SU_{\mathrm{f}}(3)}$
\begin{align}
\hat{Q} = \hat{T}_3 + \frac{\hat{Y}}{2}.  
\label{eq:chargeop}
\end{align}
Sandwiching the charge operator $\hat{Q}$ between the collective
baryon wave functions, we get the charge of the light-quark pair
inside the corresponding baryon. In order to yield the correct charge 
of the baryon concerned, we have to introduce in addition the charge
of a heavy quark inside it, as mentioned previously already. Thus, 
the $Q^2$ dependence of the electric form factor of a heavy baryon in
the present scheme is solely governed by the light quarks. The
contribution of the pointlike heavy quark is just its own constant
charge $e_Q$ as given in Eq.~\eqref{eq:app1}. Though this mean-field 
approximation may be a crude one, the $Q^2$ dependence of the electric
form factors will explain a certain characteristics of the electric
structure of the heavy baryons. 

\begin{figure}[htp]
\centering
\includegraphics[scale=0.226]{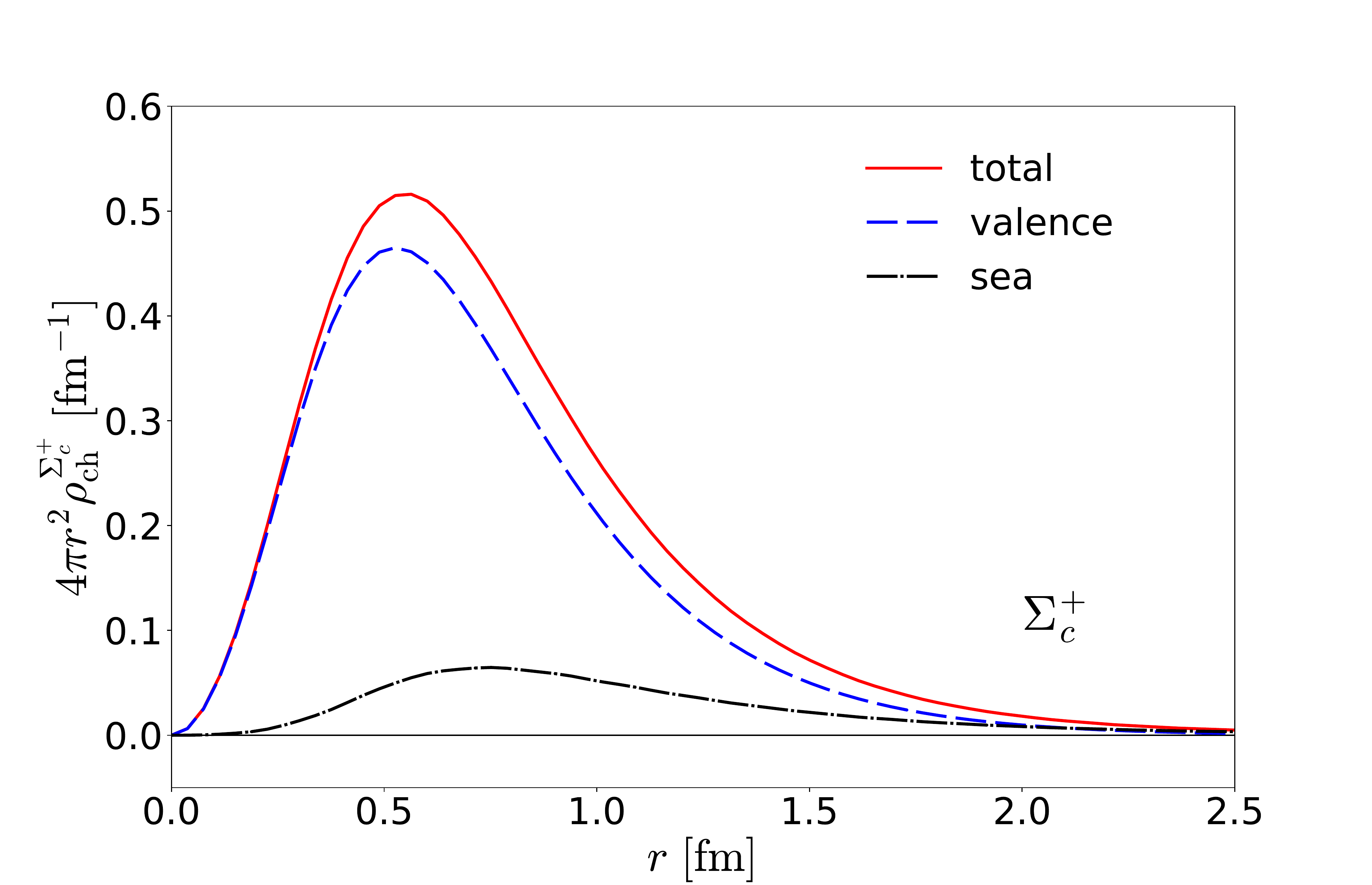}
\includegraphics[scale=0.226]{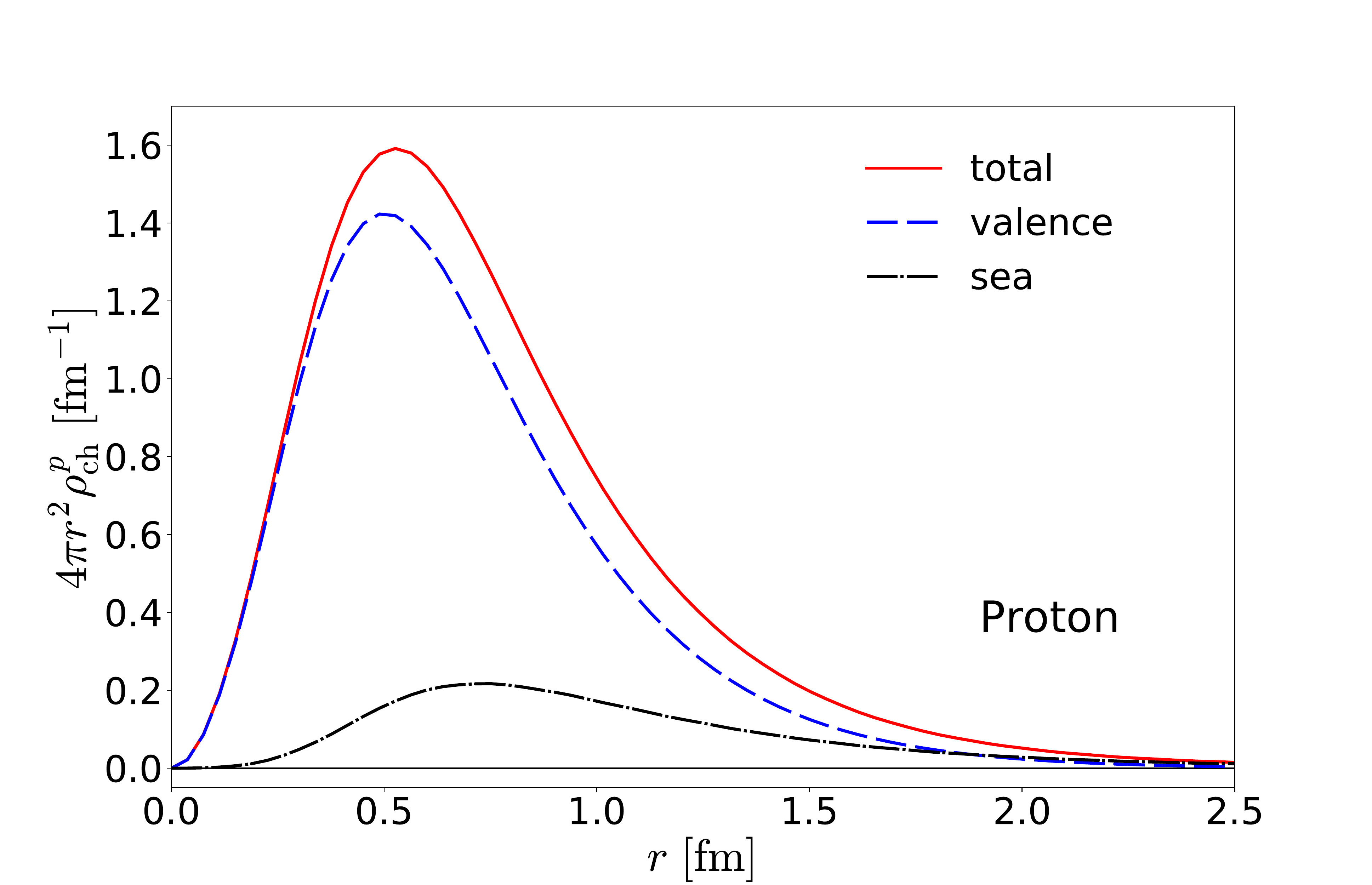}
\caption{Electric charge densities of the soliton ($J=1)$ for the  
$\Sigma_c^+$ soliton and the proton charge densities. In the left
panel, the electric charge density of $\Sigma_c^+$ is drawn whereas in
the right panel that of the proton is depicted. The dashed curve
represents the contribution of the valence quarks, while the
dot-dashed one illustrates that of the sea quarks. The solid curve
shows the total density.} 
\label{fig:1}
\end{figure}
In the present mean-field approach, the light-quark dynamics inside
both a heavy baryon and a light baryon is treated on the same
footing. Only difference lies in the different $N_c$ counting
factor, as explained previously. Thus, it is of great interest to
examine the electric charge and magnetic densities of the heavy
baryons with those of the proton and the neutron, before we compute
the EM form factors of the heavy baryons. In the left panel of
Fig.~\ref{fig:1}, we draw the electric charge densities of the soliton
for $\Sigma_c^+$, which consists of the light-quark pair ($ud$) with
spin $J=1$. Note that $\Sigma_c^+$ is a positive-charged member of the 
baryon antitriplet. The heavy quark inside $\Sigma_c^+$ is assumed
to be located at rest at the center of it. So, its charge density is
just given by the delta function. The results are compared with those
of the proton depicted in the right panel of Fig.~\ref{fig:1}. The 
general feature of the charge densities of the soliton of the
light-quark pair inside $\Sigma_c^+$ is almost the same as the proton
one. The electric charge of the light-quark pair inside $\Sigma_c^+$
is $+1/3$ whereas the proton has $+1$. Thus, both the electric charge 
densities are positive definite over the whole $r$ region. The 
difference between these two electric charge densities is found only
in the strength of the electric charge. Hence, the proton electric
charge density turns out to be approximately three times larger than
that of the soliton for $\Sigma_c^+$. The sea-quark polarizations
show marginal effects on both the electric charge densities.  

\begin{figure}[htp]
\centering
\includegraphics[scale=0.226]{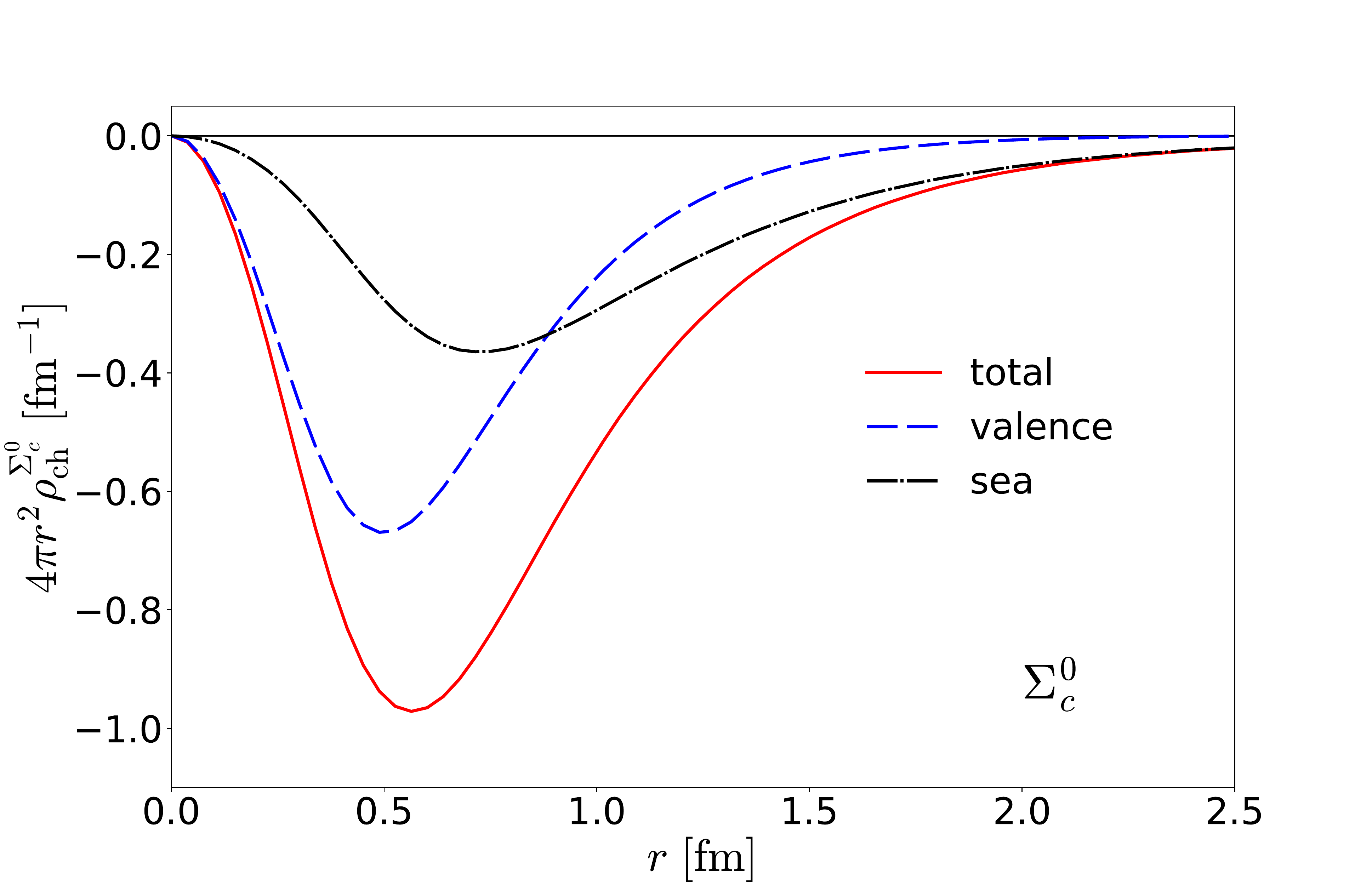}
\includegraphics[scale=0.226]{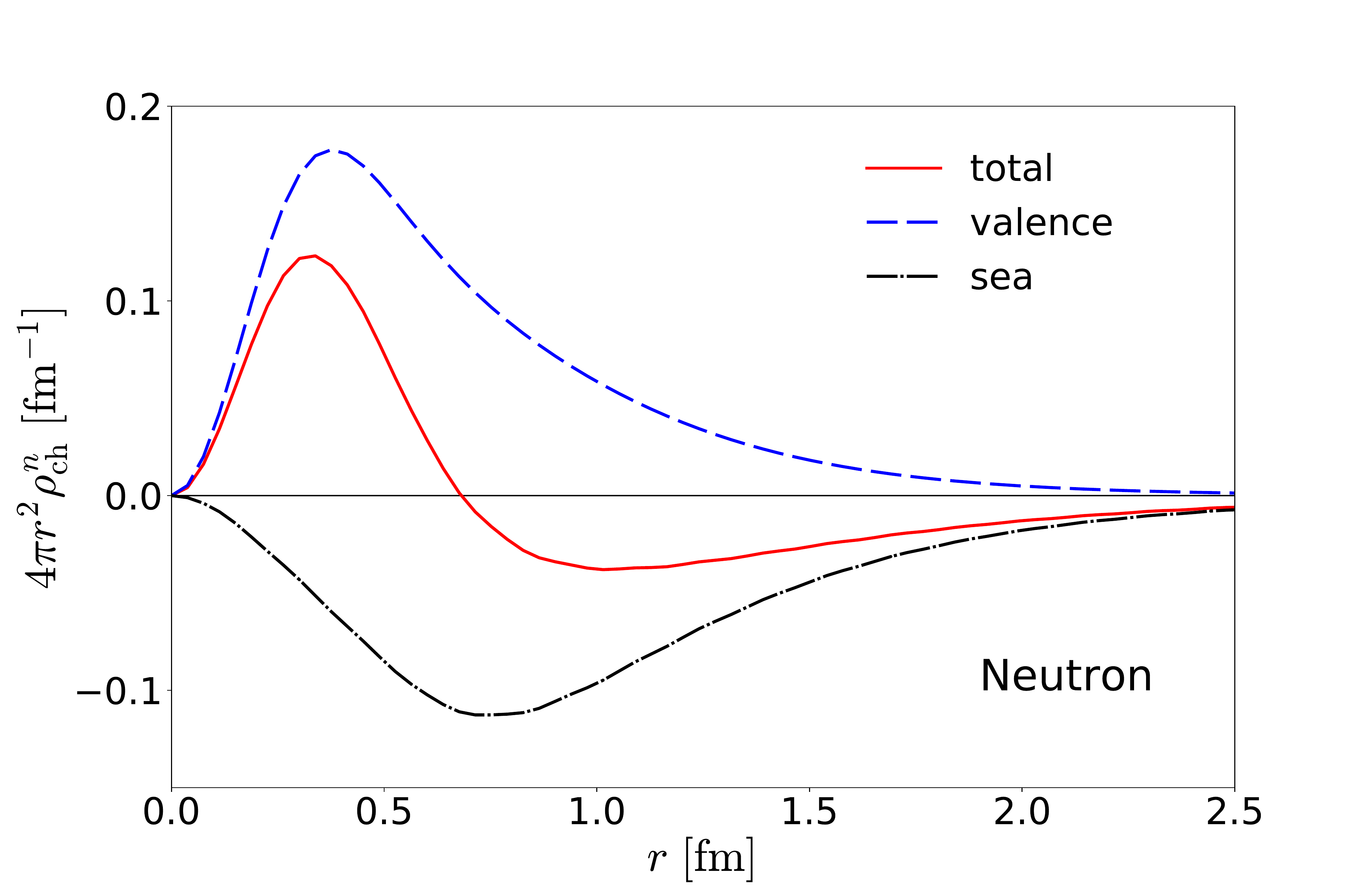}
\caption{Electric charge densities of the soliton ($J=1)$ for 
$\Sigma_c^0$ and the neutron charge densities. In the left panel,
charge density of $\Sigma_c^0$ is drawn whereas in the right panel
that of the neutron is depicted. The dashed curve represents the
contribution of the valence quarks, while the dot-dashed one
illustrates that of the sea quarks. The solid curve shows the total
contribution.} 
\label{fig:2}
\end{figure}
Figure~\ref{fig:2} compares the electric charge densities of the
soliton inside the neutral member $\Sigma_c^0$ of the baryon
antitriplet with those of the neutron ones. Since the $\Sigma_c^0$
contains two down valence quarks, the charge distribution of the
soliton inside $\Sigma_c^0$ becomes negative. Apart from the sign of the 
densities, the general behavior of the light-quark electric charge
density of $\Sigma_c^0$ is very similar to that of the proton or 
$\Sigma_c^+$. The sea-quark polarization inside $\Sigma_c^0$ is a
stronger than those inside the $\Sigma_c^+$ or proton case. On the
other hand, the neutron electric charge density is rather different
from that of $\Sigma_c^0$. In this case, the valence quarks govern the
inner part of the neutron density, whereas the sea-quark polarization
is dominant over its tail part. Thus even though the $\Sigma_c^0$ is
the neutral baryon, its light-quark charge density behave very
differently, compared with the neutron density. We will soon see that
this difference will be clearly shown in the electric form factors of
the neutral heavy baryons. All other charge densities of the
positive-charged heavy baryons are very similar to that of
$\Sigma_c^+$, and those of the neutral ones to that of $\Sigma_c^0$.

\begin{figure}[htp]
\centering
\includegraphics[scale=0.3]{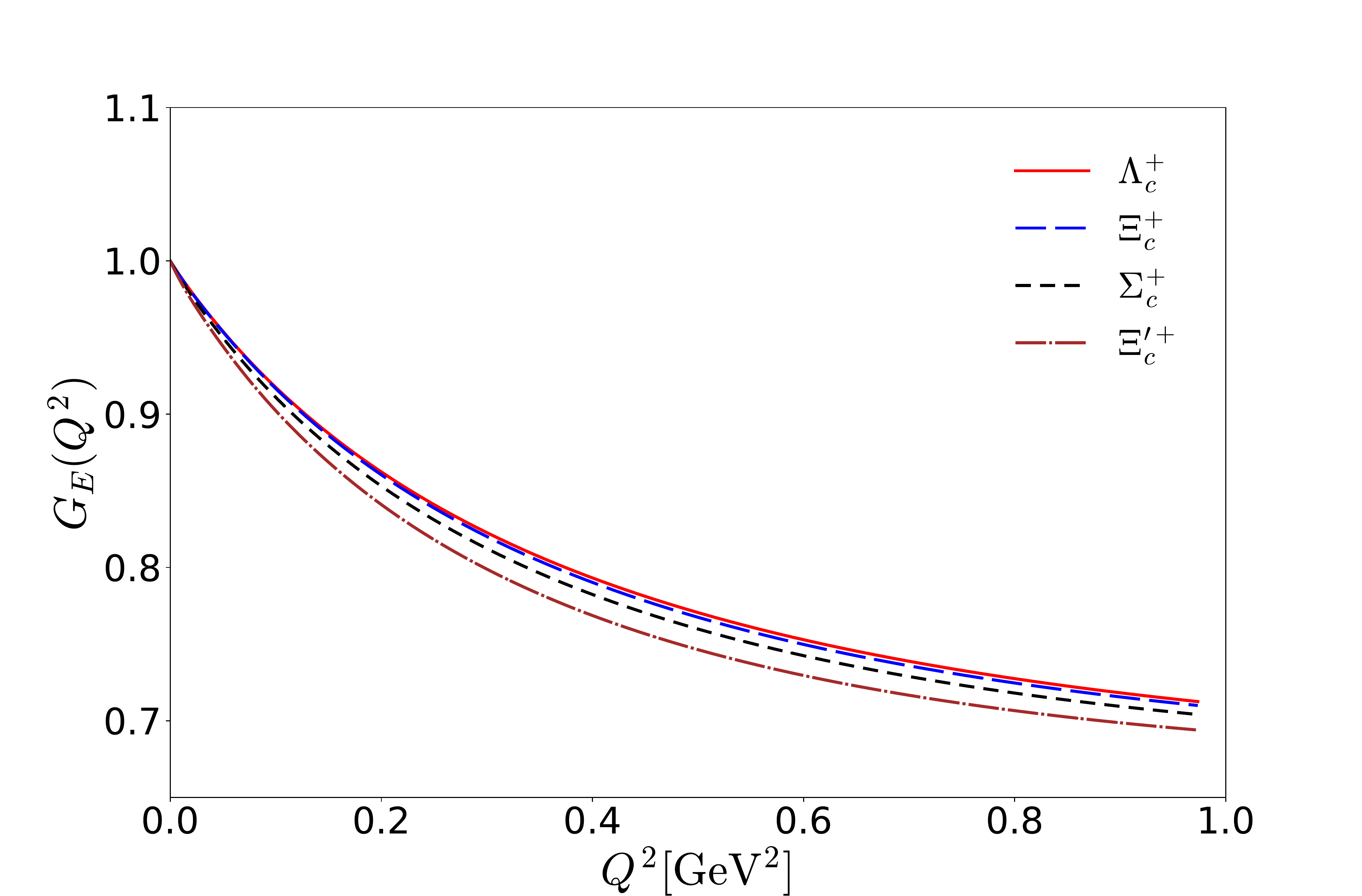}
\caption{Electric form factors of the singly positive-charged charmed  
baryons. The solid curve draws that of $\Lambda_c^+$, the
long-dashed one for $\Xi_c^+$, the dashed one for $\Sigma_c^+$, and 
the dot-dashed one for $\Xi_c^{\prime +}$.} 
\label{fig:3}
\end{figure}

\begin{figure}[htp]
\centering
\includegraphics[scale=0.3]{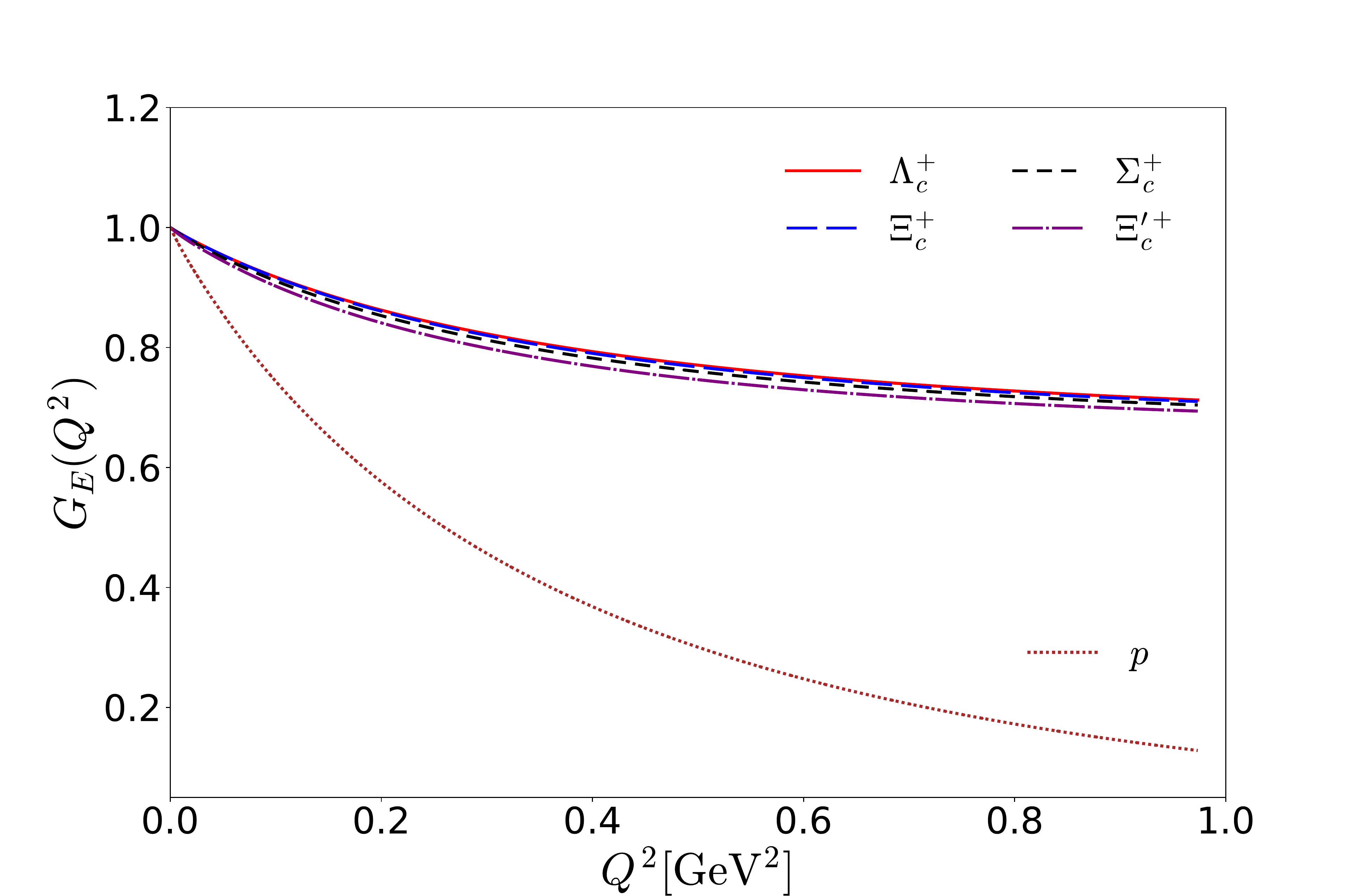}
\caption{Electric form factors of the singly positive-charged 
charmed baryons in comparison with that of the proton. The solid curve
draws that of $\Lambda_c^+$, the long-dashed one for $\Xi_c^+$, the
dashed one for $\Sigma_c^+$, and the dot-dashed one for
$\Xi_c^{\prime +}$. The short-dashed curve depicts the proton
electric form factor.}  
\label{fig:4}
\end{figure}
In Fig.~\ref{fig:3}, the electric form factors of the singly
positive-charged heavy baryons with spin $1/2$ are drawn as functions
of $Q^2$. They decrease monotonically as $Q^2$ increases. This feature
is very similar to that of the proton, which is already expected from
the comparison of the charge densities in Fig.~\ref{fig:1}. So, it is
also of great interest to compare the results of Fig.~\ref{fig:3} with
the proton electric form factor, as shown in Fig.~\ref{fig:4}. The 
electric form factor of the proton was obtained within the same
framework with exactly the same parameters. The comparison exhibits a
remarkable difference. The electric form factor of the proton falls
off much faster than those of the singly positive-charged heavy
baryons. It reveals a profound physical meaning: The heavy baryons are
electrically compact objects, so that they are much smaller than the
proton. This will be more clearly seen in the results of the charge
radii which will be discussed later.  

\begin{figure}[htp]
\centering
\includegraphics[scale=0.3]{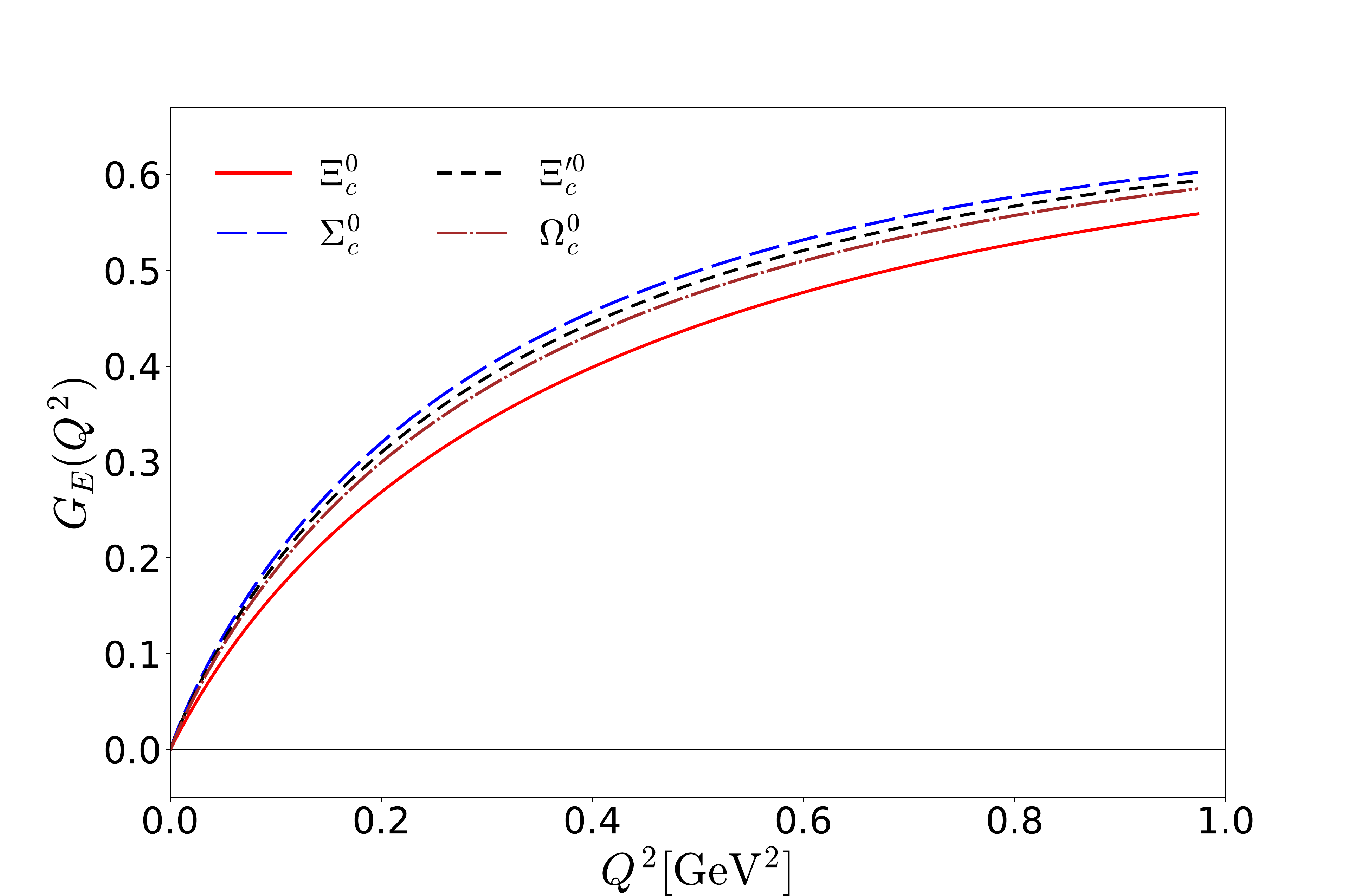}
\caption{Electric form factors of the neutral charmed
baryons. The solid curve draws that of $\Xi_c^0$, the
long-dashed one for $\Sigma_c^0$, the dashed one for 
$\Xi_c^{\prime 0}$, and the dot-dashed one for $\Omega_c^{0}$.} 
\label{fig:5}
\end{figure}
\begin{figure}[htp]
\centering
\includegraphics[scale=0.3]{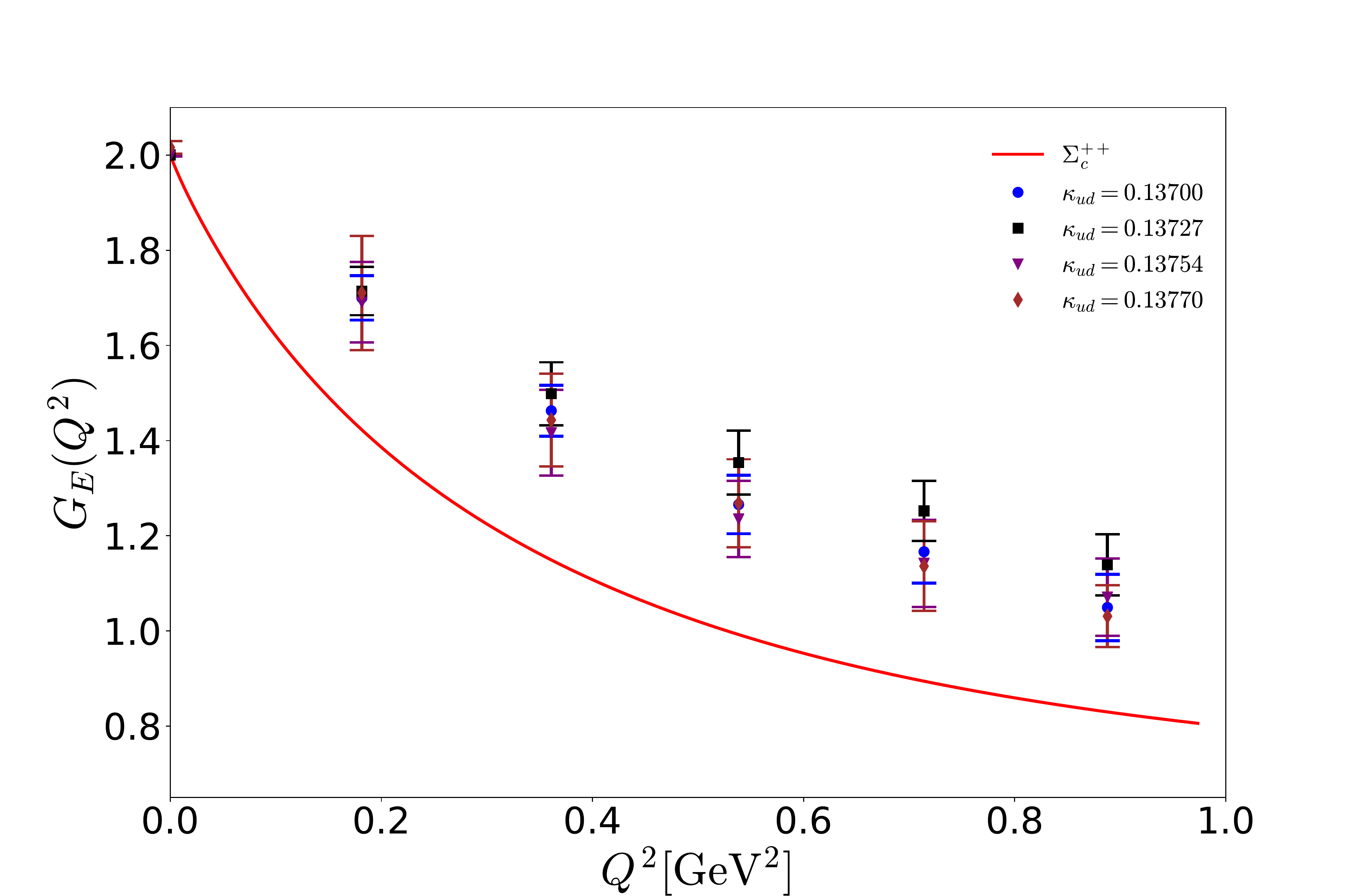}
\caption{Electric form factor of $\Sigma_c^{++}$ as a function of
  $Q^2$. The result is compared with those from lattice
  QCD~\cite{Can:2013tna}. $k_{ud}$ denotes the light-quark hopping
  parameter~\cite{Can:2013tna}.}    
\label{fig:6}
\end{figure}
In Fig.~\ref{fig:5}, we show the results of the electric form factors
of the neutral heavy baryons. They start to rise fast 
and then slow down as $Q^2$ increases. The results are rather different
from that of the neutron as discussed already in Fig.~\ref{fig:2}. The
neutron electric form factor increases also as $Q^2$ increases up to
around $0.4\,\mathrm{GeV}^2$ and then starts to decrease very
slowly~\cite{Kim:1995mr, Praszalowicz:1998jm}. The experimental data
also confirm this behavior of the neutron form
factor~\cite{Sulkosky:2017prr}. As shown in Fig.~\ref{fig:2}, the
neutron charge density is governed by the up quarks in the inner part
of the neutron, whereas the negative-charged down quark dominates its
tail part. On the other hand, the charge densities of the neutral
heavy baryons are rather similar to those of the positive-charged ones
except for the sign. Accordingly, the electric form factors of the
neutral heavy baryons increase slowly and monotonically as $Q^2$
increases. However, one should bear in mind that the present
mean-field approach is only valid in the lower $Q^2$ region, say, up
to around $1\,\mathrm{GeV}^2$ or even lower values of $Q^2$. 

In Fig.~\ref{fig:6}, we draw the numerical result of the electric form
factor of $\Sigma_c^{++}$, comparing it with those from lattice
QCD~~\cite{Can:2013tna}. Figure~\ref{fig:6} shows that the present
result falls off faster than the lattice ones. However, we want to
emphasize on the fact that the lattice data on the electric form
factor of the proton with the unphysical value of the pion mass tend
to decrease slower than the experimental data. For example, all the
lattice calculations~\cite{Capitani:2015sba, Abdel-Rehim:2015jna,
Djukanovic:2015hnh, Chambers:2017tuf} yield the results of the nucleon
electric form factor, which fall off rather slowly in comparison with
the experimental data. Even a very recent lattice calculation at the 
physical point~\cite{Alexandrou:2017ypw} shows a similar feature. The
same tendency was also found in the case of the tensor and anomalous  
tensor form factors of the nucleon~\cite{Ledwig:2010tu,
  Ledwig:2010zq}. The lattice results of these form 
factors~\cite{Gockeler:2005cj, Gockeler:2006zu} also fall off much
more slowly than those results of the $\chi$QSM.

\begin{figure}[htp]
\centering
\includegraphics[scale=0.226]{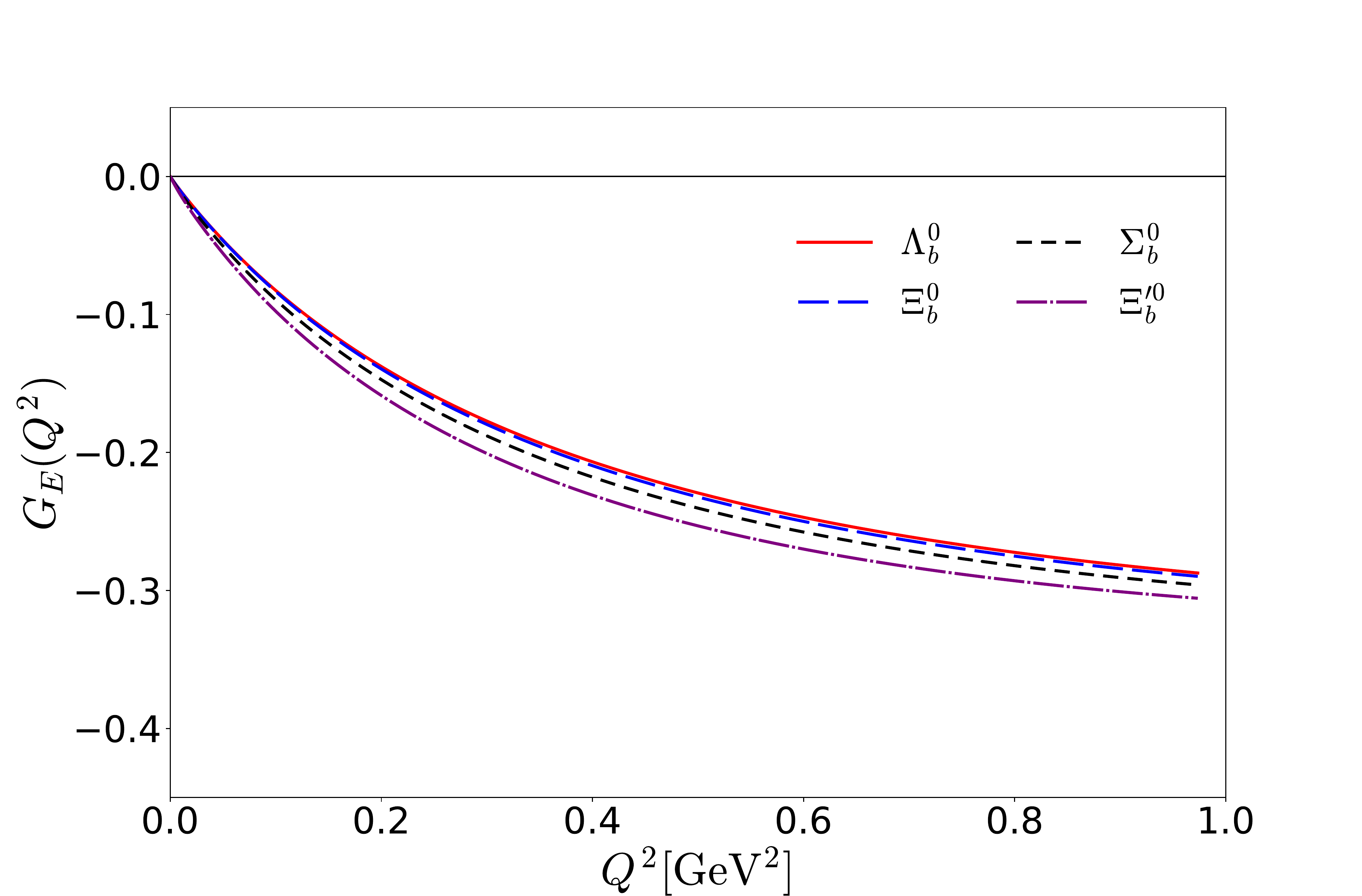}
\includegraphics[scale=0.226]{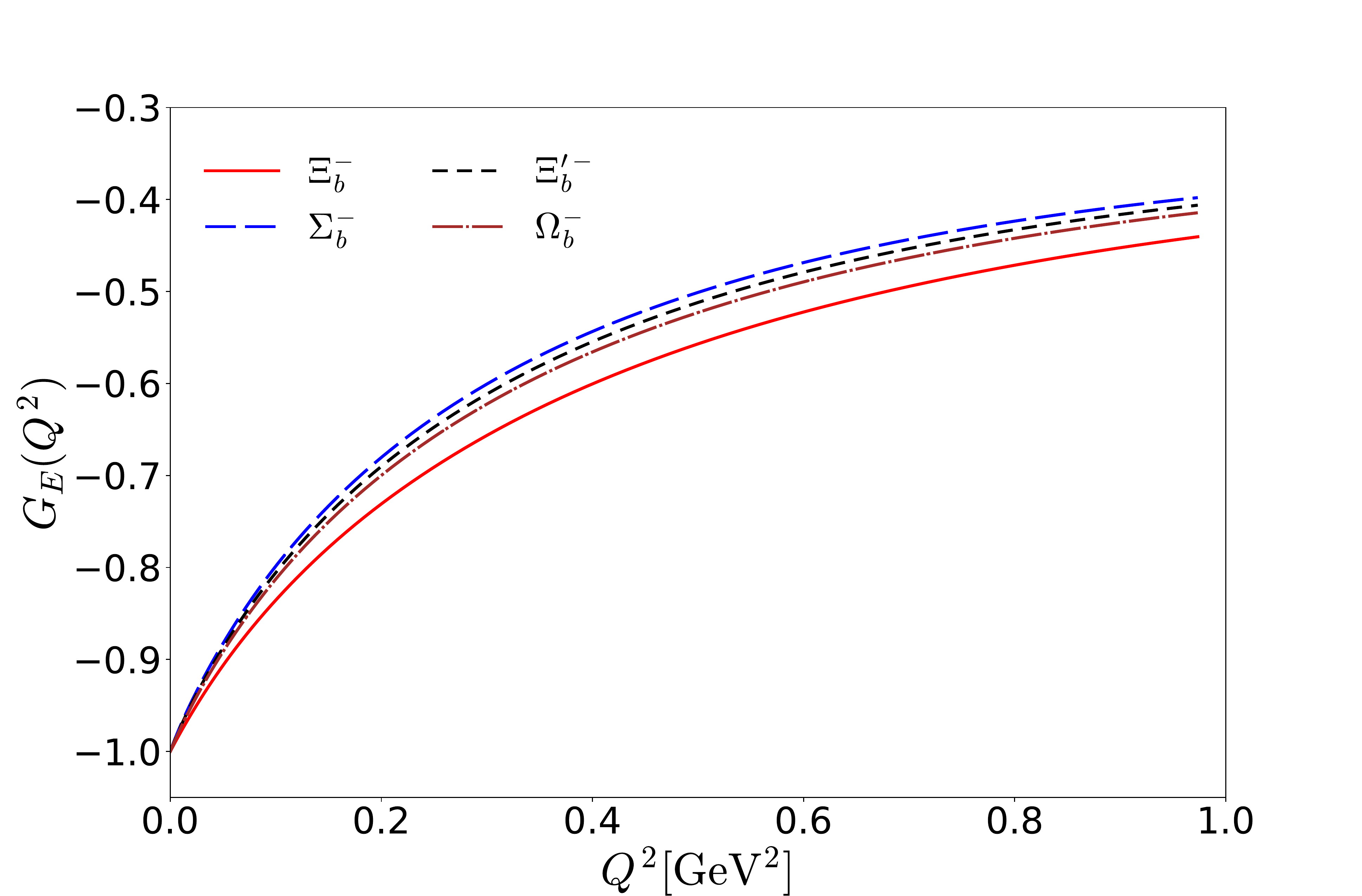}
\includegraphics[scale=0.226]{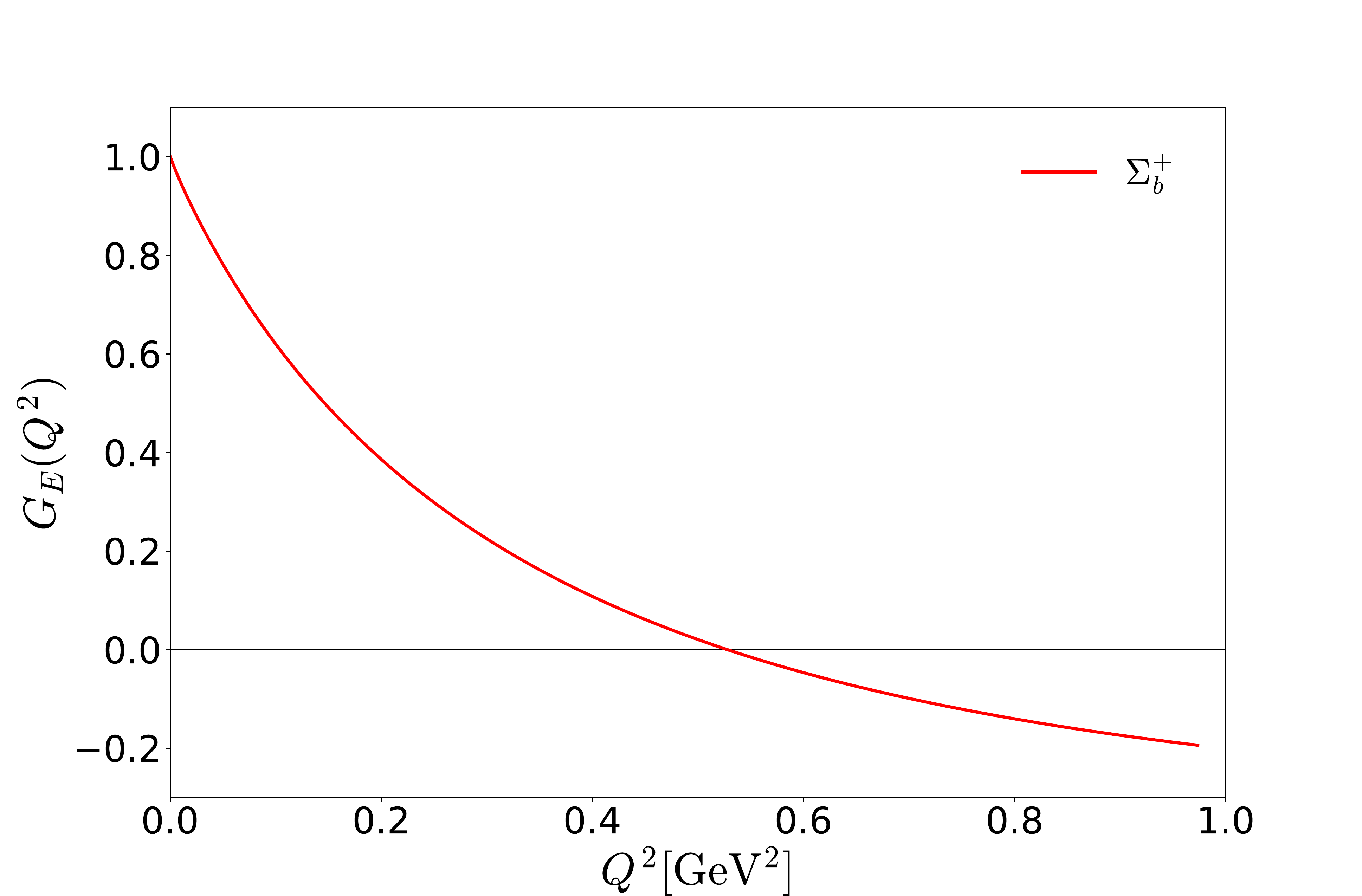}
\caption{Electric form factors of the bottom baryons as functions of
  $Q^2$. In the upper left panel, those of the neutral bottom baryons
  are drawn. The upper right panel presents those of the
  negative-charged bottom baryons. In the lower panel, that of
  $\Sigma_b^+$ is drawn.}  
\label{fig:7}
\end{figure}
In the present mean-field approach, there is in principle no difference
between the electric form factors of the charmed baryons and those of
the bottom baryons, since the same light quarks govern the $Q^2$
dependence of the form factors. The charge of the bottom quark makes
their electric form factors distinguished from those of the charmed
baryons. In Fig.~\ref{fig:7}, we draw the electric form factors of the
bottom baryons. In the upper left panel of Fig.~\ref{fig:7}, those of
the neutral bottom baryons are depicted. Interestingly, they are all 
negative, which are different from those of the neutral charmed
baryons. This can be understood from the differrent charges of the
charm and bottom quarks. The upper right panel of Fig.~\ref{fig:7}
present the electric form factors of the negative-charged bottom
baryons. That of $\Sigma_b^+$ is illustrated in the lower panel of
Fig.~\ref{fig:7}. We find that it becomes negative at around
$0.55\,\mathrm{GeV}^2$, which is again due to the negative charge
$e_b=-1/3$ of the bottom quark.          

\begin{table}[htp]
  \centering
\caption{Electric charge radii of the charmed baryons in units of 
  $\mathrm{fm}^2$.} 

\begin{tabular}{ c  c c c } 
 \hline 
 \hline 
Baryon &  $\langle r^{2} \rangle_{E}^{B_c  (m_{s} = 0 \, \mathrm{MeV})}
         $  &  $\langle r^{2} \rangle_{E}^{B_c (m_{s} = 174 \,\mathrm{MeV})}  $ 
         & \cite{Can:2013tna} \\ 
 \hline
 $\Lambda_{c}^{+}$&0.26&0.24  &  --\\
$\Xi_{c}^{+}$&0.26&0.24 &  --\\
$\Xi_{c}^{0}$&-$0.52$&-$0.51$ &  --\\
 \hline
$\Sigma_{c}^{++}$&0.60&0.59 &  $0.234 \pm 0.037$\\
$\Sigma_{c}^{+}$&0.30&0.27  &  --\\
$\Sigma_{c}^{0}$&-$0.60$&-$0.66$ &  --\\
$\Xi_{c}^{\prime +}$&0.30&0.30   &  --\\
$\Xi_{c}^{\prime 0}$&-$0.60$&-$0.63$  &  --\\
$\Omega_{c}^{0}$&-$0.60$&-$0.60$ &  --\\
 \hline 
 \hline
\end{tabular}
\label{tab:1}
\end{table}
\begin{table}
\caption{Electric charge radii of the bottom baryons in units of
  $\mathrm{fm}^2$.}
  \centering
\begin{tabular}{  c  c  c  } 
 \hline 
 \hline 
Baryon &  $\langle r^{2} \rangle_{E}^{B_b(m_{s} = 0 \, \mathrm{MeV})} $ 
 &  $\langle r^{2} \rangle_{E}^{B_b (m_{s} = 166 \, \mathrm{MeV})}  $ \\ 
 \hline
 $\Lambda_{b}^{0}$&0.26&0.24 \\
$\Xi_{b}^{0}$&0.26&0.24      \\
$\Xi_{b}^{-}$&$0.52$ & $0.51$      \\
 \hline
$\Sigma_{b}^{+}$&0.60&0.60   \\
$\Sigma_{b}^{0}$&0.30&0.27   \\
$\Sigma_{b}^{-}$&$0.60$ & $0.66$   \\
$\Xi_{b}^{0}$&0.30&0.30      \\
$\Xi_{b}^{-}$& $0.60$ & $0.63$      \\
$\Omega_{b}^{-}$& $0.60$ & $0.60$  \\
 \hline 
 \hline
\end{tabular}
\label{tab:2}
\end{table}
In the present approach, more important observables are the electric
charge radii, since they are determined by the behavior of the
electric form factors near $Q^2=0$ and provide information on the
sizes of the heavy baryons. The electric charge radii of the baryons
is defined by 
\begin{align}
\langle r^2 \rangle_E^{B_Q} = -\frac{6}{G_E^B(0)} \left. \frac{d
  G_E^{B}(Q^2)}{dQ^2}   \right |_{Q^2=0} .  
\label{eq:charge_radii}
\end{align}
Note that the electric charge radius given in
Eq.~\eqref{eq:charge_radii} is normalized by the value of 
the corresponding electric form factor at $Q^2=0$ for a charged heavy
baryon. As for a neutral one, we do not normalize it. Since the
heavy-quark contribution to the electric form 
factors is just the constant charge of the related heavy baryon, it
does not contribute to the electric charge radii. Thus, the electric
charge radii are solely determined by the solitons of the light-quark
pair. In Table~\ref{tab:1}, we list the results of the electric charge
radii of the charmed baryons. In the second and third columns, those
with $m_{\mathrm{s}}=0$ and $m_{\mathrm{s}}=174$ MeV are presented. 
The results show explicitly that in the $\mathrm{SU}_{\mathrm{f}}(3)$
symmetric case the $U$-spin symmetry is preserved, i.e. we have the
following relations in each representation
\begin{align}
\langle r^2\rangle_E^{\Lambda_c^+} &= \langle r^2\rangle_E^{\Xi_c^+} =
  -\frac12 \langle r^2\rangle_E^{\Xi_c^0},\cr
\langle r^2\rangle_E^{\Sigma_c^+} &= \langle
 r^2\rangle_E^{\Xi_c^{\prime +}} = -\frac12 \langle
r^2\rangle_E^{\Sigma_c^0} -\frac12 \langle r^2 
\rangle_E^{\Xi_c^{\prime 0}} = -\frac12 \langle r^2 
\rangle_E^{\Omega_c^{0}} = \frac12 \langle r^2 
\rangle_E^{\Sigma_c^{++}}.   
\end{align}
The results listed in Table~\ref{tab:1} indicate that the effects of 
$\mathrm{SU}_{\mathrm{f}}(3)$ are marginal. We see that the baryon
antitriplet have smaller sizes than those of the baryon sextet with
spin 1/2. Moreover, as we already mentioned in discussion of
Fig.~\ref{fig:2}, the electric charge radii of the positive-charged
heavy baryons are noticeably smaller than that of the proton that is
experimentally known to be $\langle r^2 \rangle_E^p=(0.70-0.74)\,
\mathrm{fm}^2$~\cite{PDG2017}.  The present result of the electric
charge radius for $\Sigma_c^{++}$ is compared with the lattice 
data~\cite{Can:2013tna}. As expected from Fig.~\ref{fig:6}, the 
present result is significantly larger than the lattice one. 
In Table~\ref{tab:2}, we list the
results of the electric charge radii of the bottom baryons. The
results are the same as those of the charmed baryons in the present
mean-field approach.  

\subsection{Magnetic form factors of the baryon sextet with spin 1/2}
We now turn our attention to the magnetic form factors of the heavy
baryons, which are expressed in Eqs.~\eqref{eq:magfinal} and
\eqref{eq:magden}. Assuming that the mass of the heavy quark is
infinitely heavy, the heavy quark does not contribute to the magnetic
form factors of the heavy baryons, since the heavy-quark contribution
is proportional to the inverse of the heavy-quark mass ($\mu_c \sim
1/m_Q$). Thus, the magnetic form factors of the heavy baryons are
completely governed by the light-quark soliton. Note that in the
present approach all the magnetic form factors of the baryon
antitriplet vanish, since the soliton inside the baryon antitriplet
has spin $J=0$. It implies that they will be ascribed to higher-order
corrections beyond the mean-field approximation and should be rather
small. Thus, in this present work, we present the results of the
magnetic form factors of the baryon sextet with spin 1/2. Those with
spin 3/2 need to be treated separately, since their spin structures
are more involved than the case of spin 1/2. The results of the spin
3/2 heavy baryons will appear elsewhere.  

\begin{figure}[htp]
\centering
\includegraphics[scale=0.226]{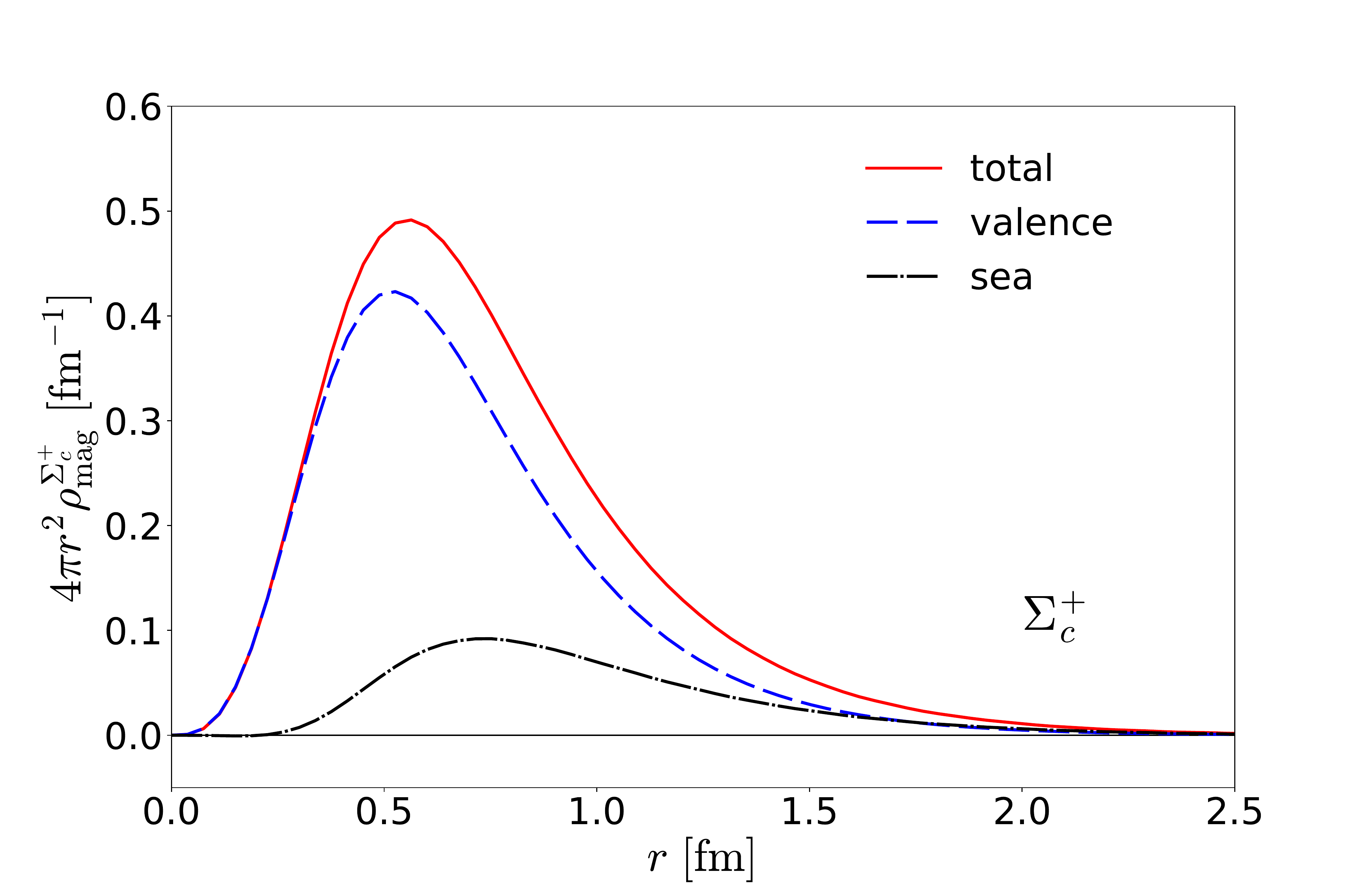}
\includegraphics[scale=0.226]{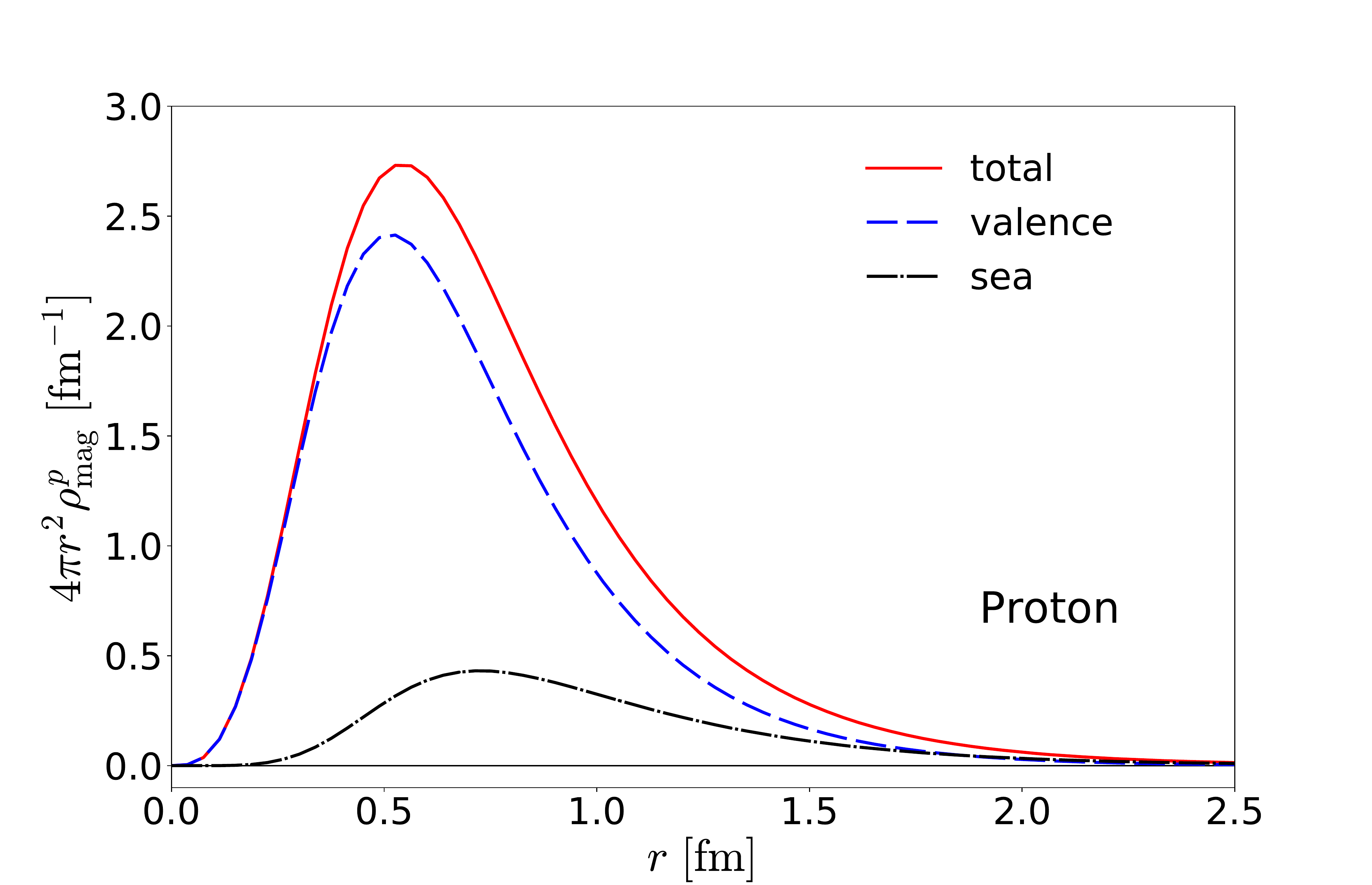}
\caption{Magnetic densities of the soliton ($J=1)$ for the
$\Sigma_c^+$ and the proton magnetic densities. In the left panel,
magnetic densities of $\Sigma_c^+$ are drawn whereas in the right panel
that of the proton are depicted. The dashed curve represents the
contribution of the valence quarks, while the dot-dashed one
illustrates that of the sea quarks. The solid curve shows the total
contribution.} 
\label{fig:8}
\end{figure}

\begin{figure}[htp]
\centering
\includegraphics[scale=0.226]{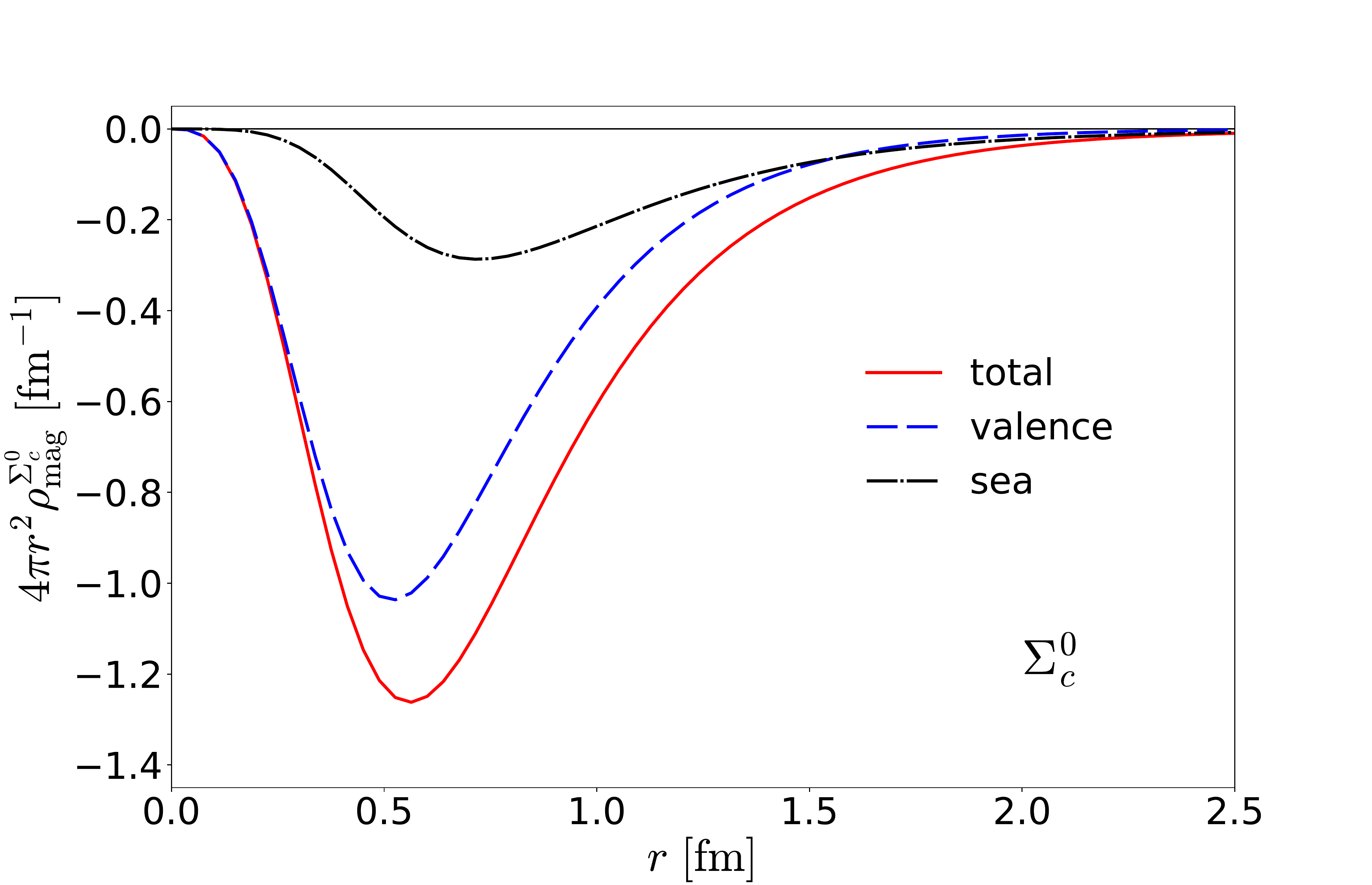}
\includegraphics[scale=0.226]{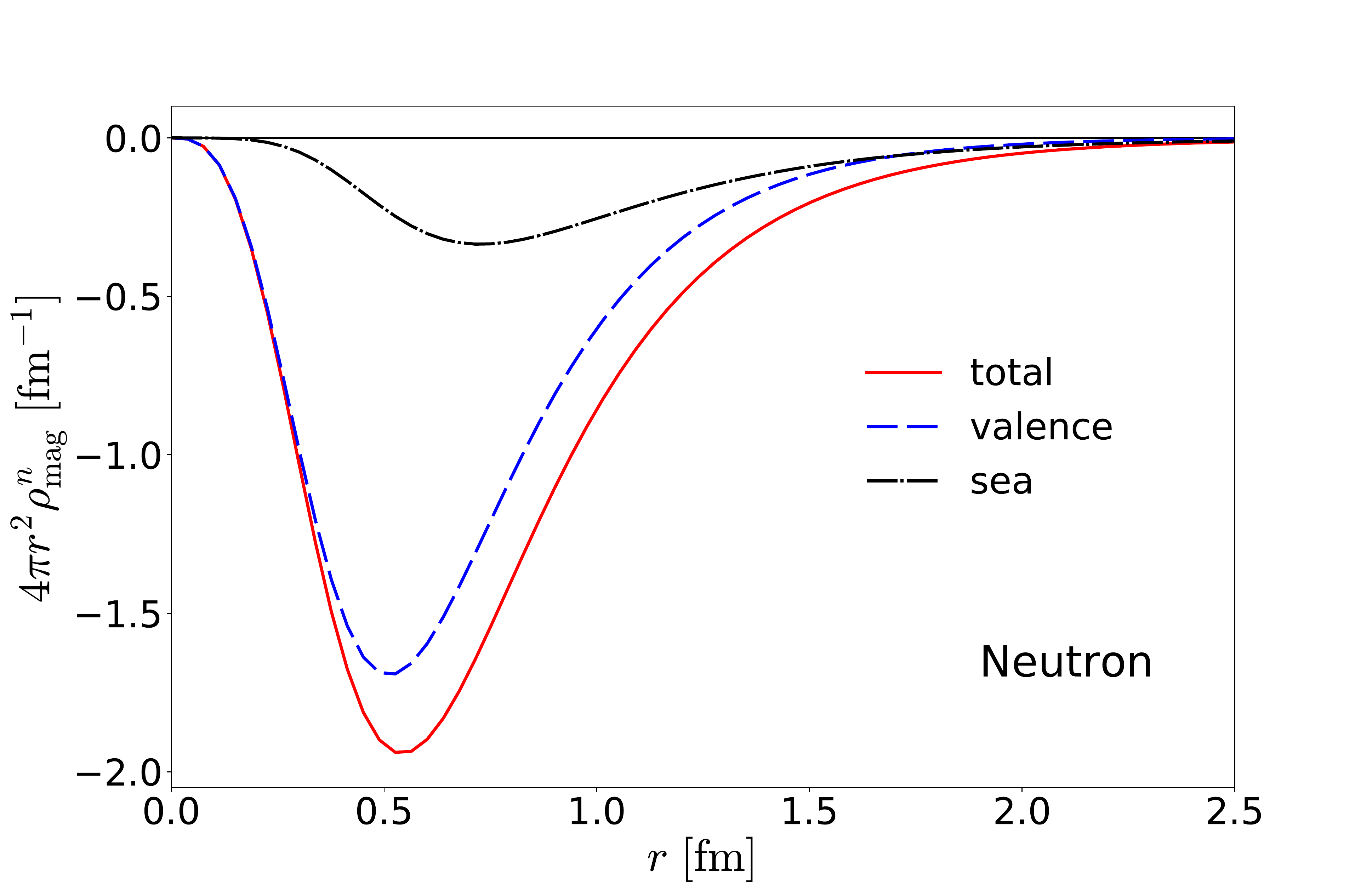}
\caption{Magnetic densities of the soliton ($J=1)$ for the
$\Sigma_c^0$ and the neutron magnetic densities. In the left panel,
magnetic density of $\Sigma_c^0$ are drawn whereas in the right panel
that of the neutron are depicted. The dashed curve represents the
contribution of the valence quarks, while the dot-dashed one
illustrates that of the sea quarks. The solid curve shows the total
contribution.} 
\label{fig:9}
\end{figure}
Figure~\ref{fig:8} compares the magnetic densities of $\Sigma_c^+$
with those of the proton. As in the case of the electric charge
densities, we find that the general feature of the $\Sigma_c^+$
magnetic densities is very similar to the proton ones. The difference
is found only in the magnitudes of the densities. In Fig.~\ref{fig:9},
the magnetic densities of $\Sigma_c^0$ are compared with those of the
neutron. Both of them look similar each other except for the
magnitudes again.

\begin{figure}[htp]
\centering
\includegraphics[scale=0.226]{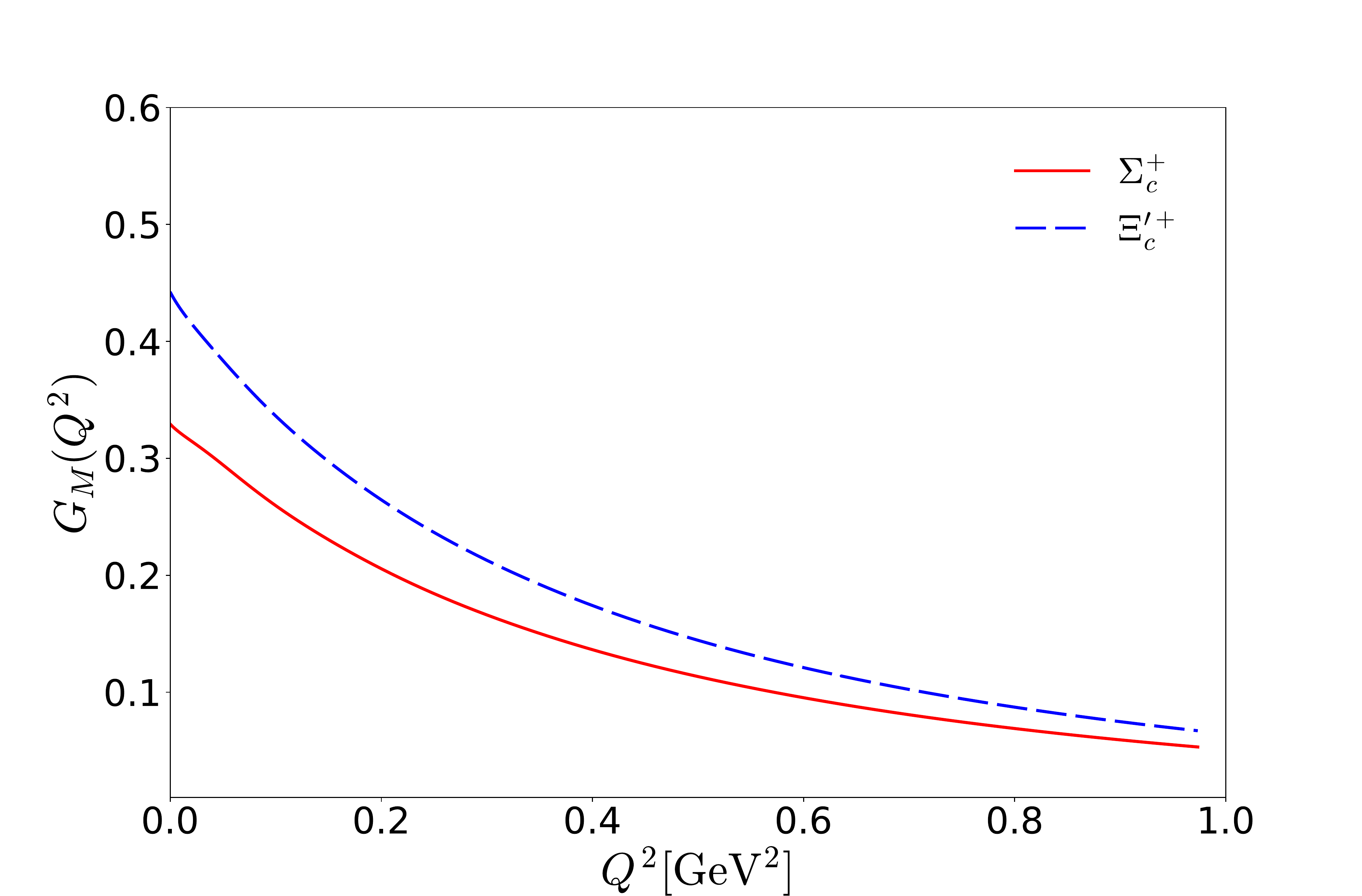}
\includegraphics[scale=0.226]{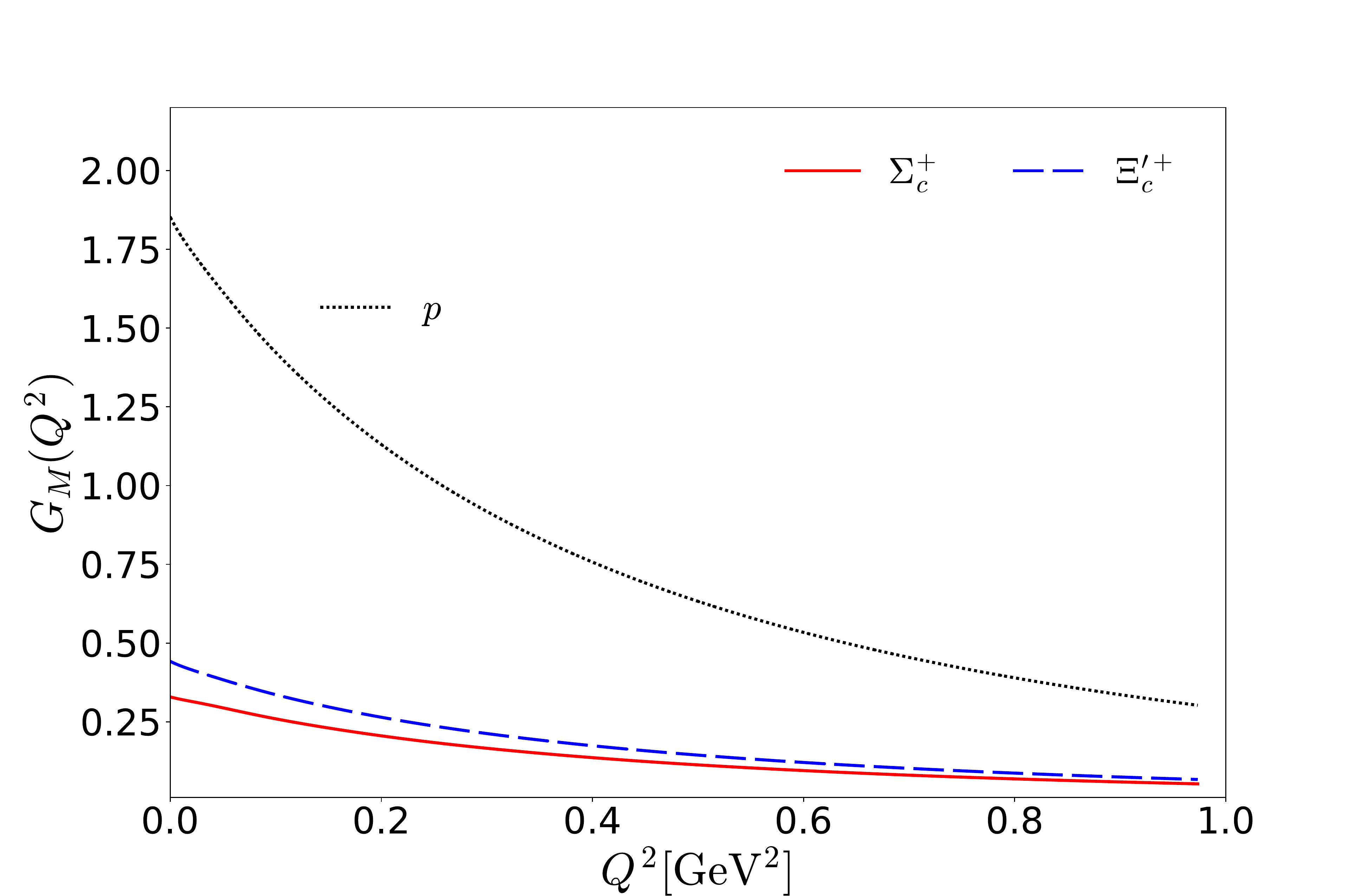}
\caption{Magnetic form factors of the singly positive-charged baryon
  sextet with $J'=1/2$. In the left panel the magnetic form factors of
  the singly positive-charged baryon sextet are drawn. The solid curve
depicts that of $\Sigma_c^+$ whereas the dashed one illustrates that
of $\Xi_c^{\prime +}$. In the right panel, the magnetic form factors
of $\Sigma_c^+$ and $\Xi_c^{\prime +}$ are compared with that of the
proton.}  
\label{fig:10}
\end{figure}
\begin{figure}[htp]
\centering
\includegraphics[scale=0.226]{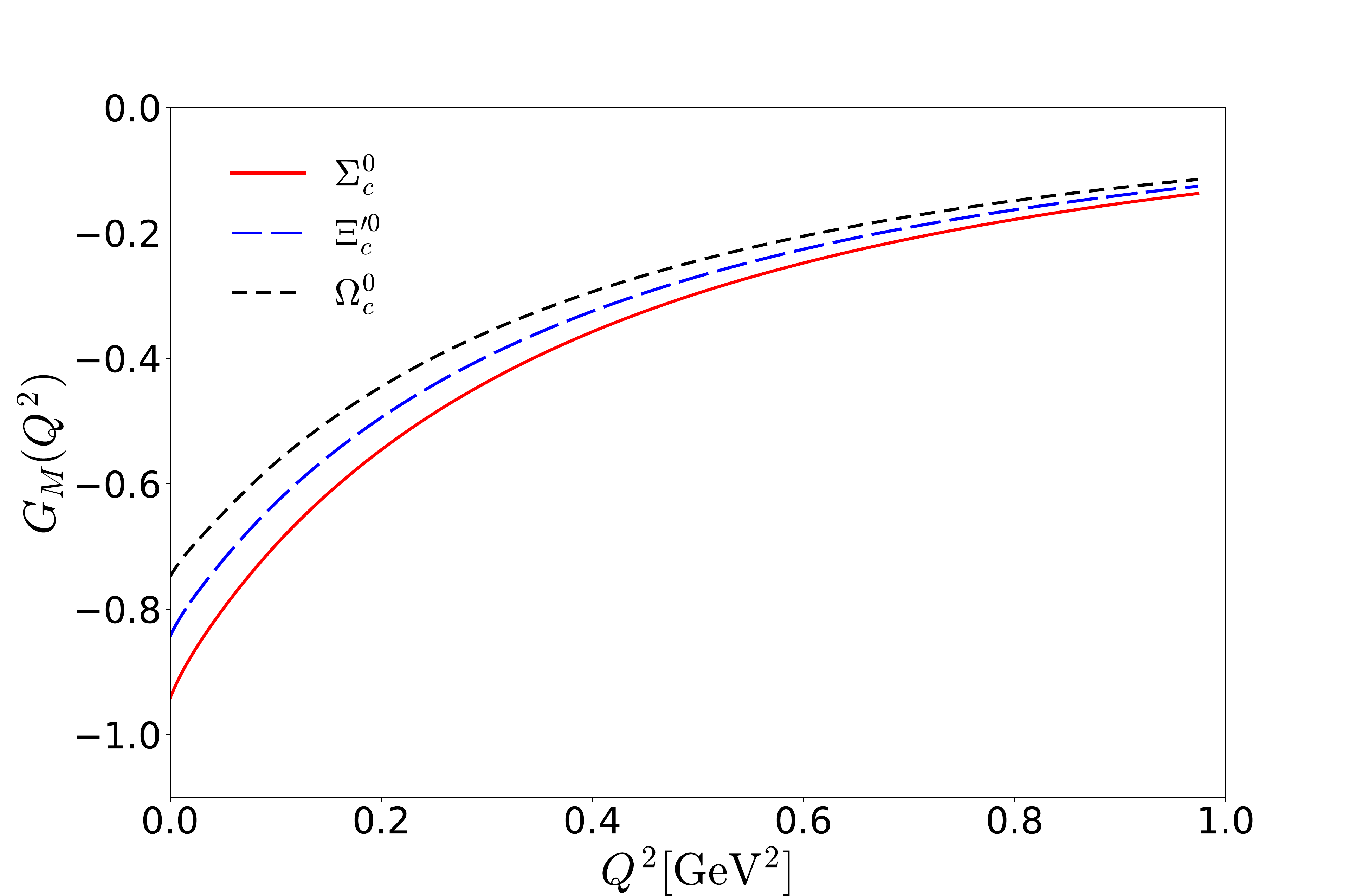}
\includegraphics[scale=0.226]{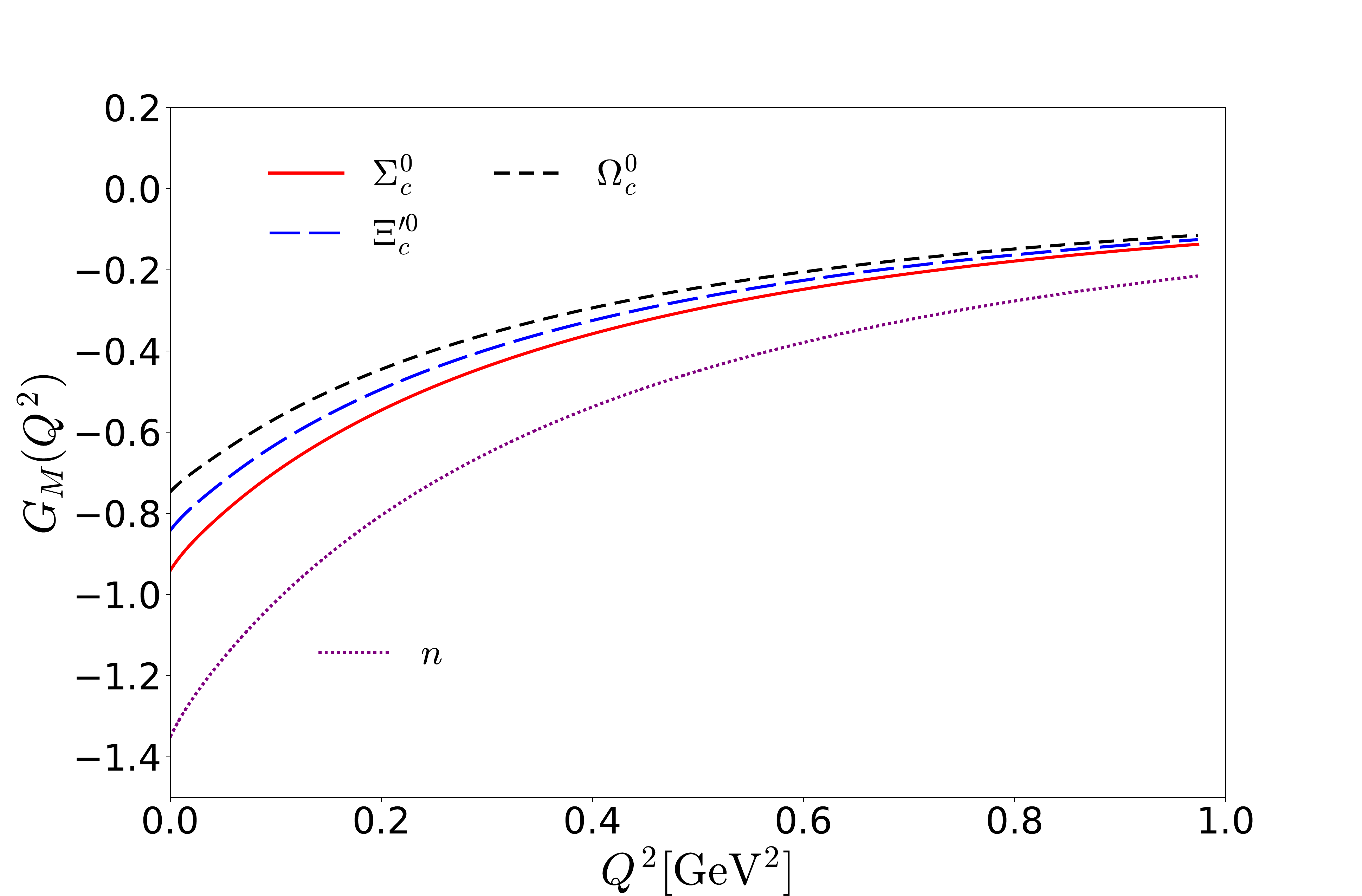}
\caption{Magnetic form factor of the neutral baryon sextet with
  $J'=1/2$. In the left panel the magnetic form factors of
  the neutral baryon sextet are drawn. The solid curve
depicts that of $\Sigma_c^0$ whereas the dashed and short dashed ones
illustrate those of $\Xi_c^{\prime 0}$ and $\Omega_c^0$,
respectively. In the right panel, the magnetic form factors 
of the neutral baryon sextet are compared with that of the
neutron.} 
\label{fig:11}
\end{figure}
\begin{figure}[htp]
\centering
\includegraphics[scale=0.3]{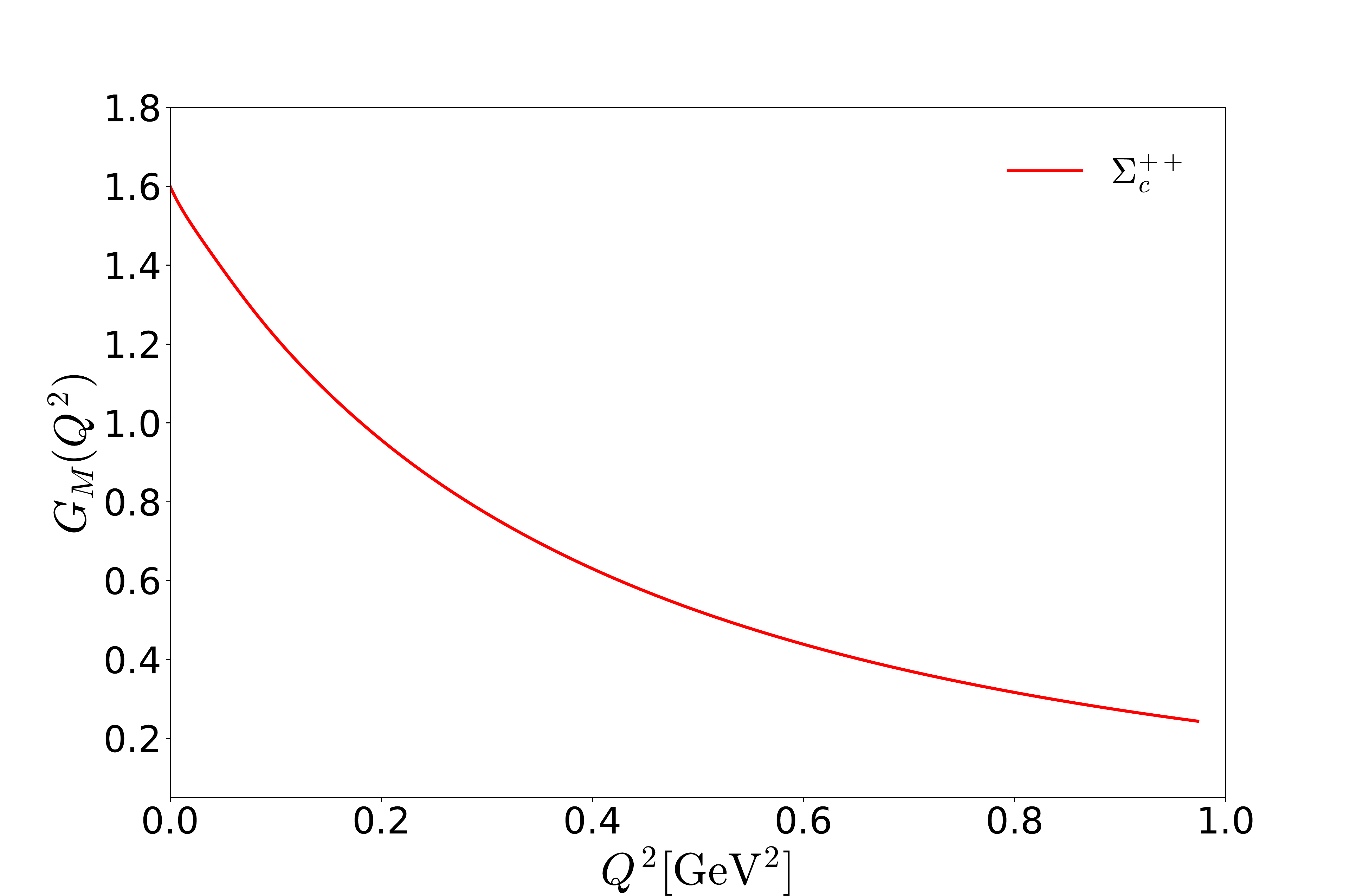}
\caption{Magnetic form factor of $\Sigma_c^{++}$.}
\label{fig:12}
\end{figure}
\begin{figure}[htp]
\centering
\includegraphics[scale=0.226]{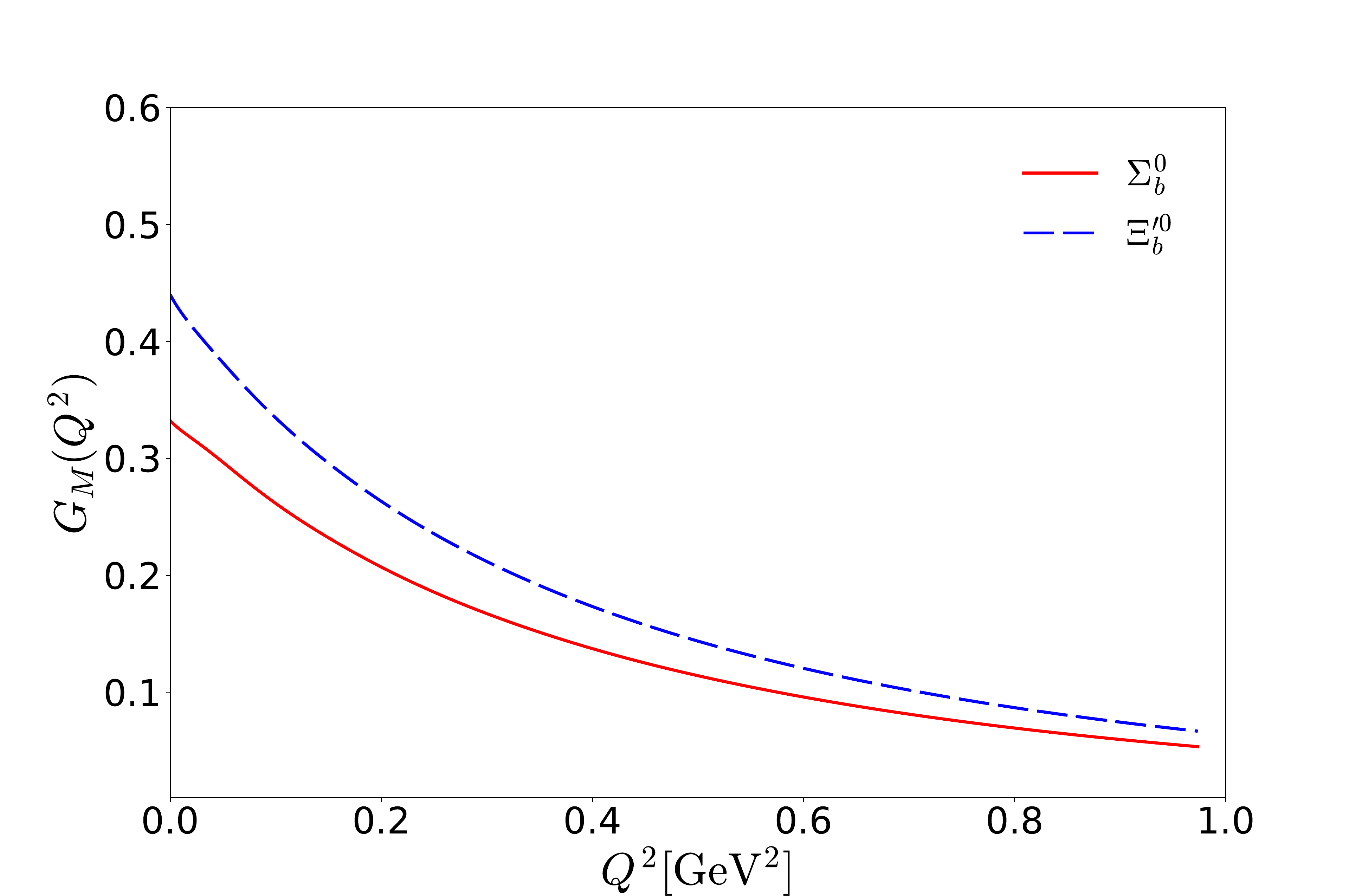}
\includegraphics[scale=0.226]{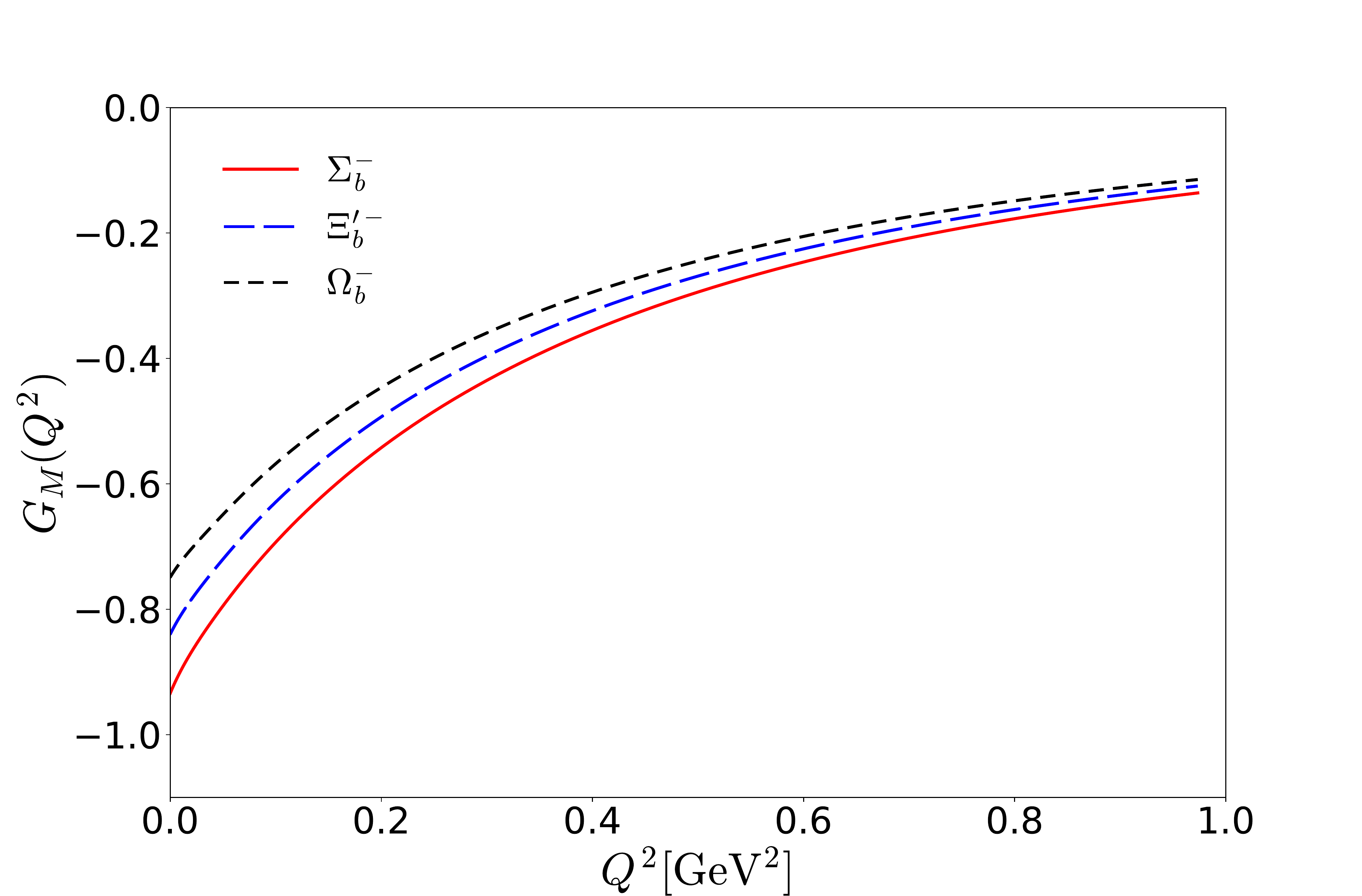}
\includegraphics[scale=0.226]{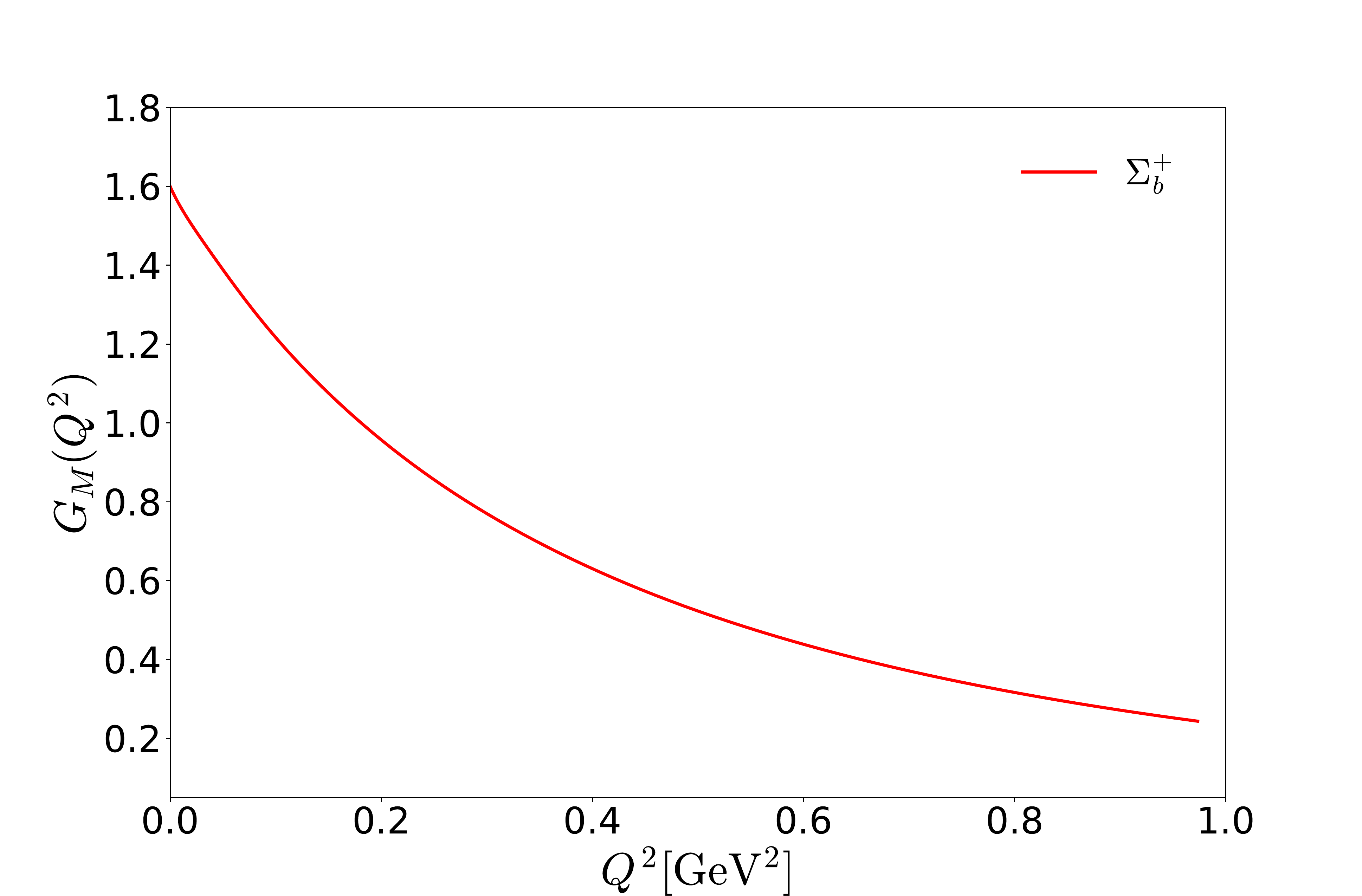}
\caption{Magnetic form factors of the bottom baryon sextet with $J'=1/2$.} 
\label{fig:13}
\end{figure}
In the left panel of Fig.~\ref{fig:10}, we show the results of the
magnetic form factors of the positive-charged baryon sextet with spin
1/2. The magnetic form factor of $\Xi_c^{\prime +}$ is larger than
that of $\Sigma_c^+$. Both form factors decrease monotonically as
$Q^2$ increases. In the right panel of  Fig.~\ref{fig:10}, we compare
the results of the magnetic form factors of $\Sigma_c^+$ and
$\Xi_c^{\prime +}$ with that of the proton. We already expect that the
protom magnetic form factor should be larger than those of these heavy
baryons from the comparison of the magnetic densities in
Fig.~\ref{fig:8}. In Figs.~\ref{fig:11} and \ref{fig:12}, we present
the magnetic form factors of all the other members of the baryon
sextet with spin 1/2. They start to fall off as $Q^2$ increases. 
For completeness, we draws in Fig.~\ref{fig:13} the results of the
magnetic form factors of the bottom baryon sextet with spin 1/2.

As in the case of the electric charge
radii, we also have the relation from the $U$-spin symmetry when 
$m_{\mathrm{s}}$ is set equal to zero:   
\begin{align}
\mu(\Sigma_c^0) = \mu (\Xi_c^{\prime 0}) = \mu(\Omega_c^0) =
  -2\mu(\Sigma_c^+) = - 2\mu(\Xi_c^{\prime +}).
\end{align}
As discussed in Ref.~\cite{Yang:2018uoj}, we can find the relations
arising from the isospin symmetry: 
\begin{align}
\mu(\Sigma_{c}^{++})\;-\;\mu(\Sigma_{c}^{+})
& =  
\mu(\Sigma_{c}^{+})\;-\;\mu(\Sigma_{c}^{0}),\cr
\mu(\Sigma_{c}^{0})\;-\;\mu(\Xi_{c}^{\prime0}) 
& =  
\mu(\Xi_{c}^{\prime0})\;-\;\mu(\Omega_{c}^{0}), \cr
2 [\mu(\Sigma_{c}^{+})\,-\,\mu(\Xi_{c}^{\prime0})]
& =  
\mu(\Sigma_{c}^{++})\,-\,\mu(\Omega_{c}^{0}).
\label{eq:coleman}  
\end{align}
Note that relations in Eq.~\eqref{eq:coleman} are also valid when
the $SU_{\mathrm{f}}(3)$ symmetry is broken.  Yet another interesting
relation is the sum rule of the baryon magnetic moments of the baryon
sextet with spin 1/2. If one adds all the magnetic moments of the
baryon sextet with spin 1/2 in the $SU_{\mathrm{f}}(3)$ symmetric
case, then we obtain the sum rule 
\begin{align}
\sum_{B_c\in \mathrm{sextet}} \mu(B_c) = 0.
\label{eq:sum}  
\end{align}
In Ref.~\cite{Kim:1997ip}, one finds a very similar relation for 
the magnetic moments of the baryon decuplet.  However, while
the sum of all the magnetic moments of the baryon decuplet is the same
as that of all the electric charges of the corresponding
baryons~\cite{Kim:1997ip}, Eq.~\eqref{eq:sum} is identical to the sum
of $2\mathcal{Q}-1$ for all the members of the 
baryon sextet, where $\mathcal{Q}$ denotes the charge of the
corresponding heavy baryon in the sextet. Thus, Eq.~\eqref{eq:sum} is
satisfied. Note that Eq.~(\ref{eq:sum}) is no more valid when the
effects of  $SU_{\mathrm{f}}(3)$ symmetry breaking are considered. 

\begin{table}[htp]
  \centering
\caption{Magnetic moments of the charmed baryon sextet with spin 1/2
  in comparison with various models. The results are given in units of
  the nuclear magneton $\mu_N$.} 
\begin{tabular}{ c  c c  |  c  c c c c c} 
 \hline 
  \hline 
Baryon & $\mu_{B_c}^{(m_{\mathrm{s}}=0 \,\mathrm{MeV})}$  &
$\mu_{B_c}^{(m_{\mathrm{s}}=174\,\mathrm{MeV})}$ & 
\cite{Yang:2018uoj}
& \cite{Oh:1991ws}  
& \cite{Scholl:2003ip}
& \cite{Faessler:2006ft} 
& \cite{Wang:2018gpl}
& \cite{Can:2013tna, Bahtiyar:2016dom}  \\   
 \hline
$\Sigma_{c}^{++}$ & 1.58   &1.60& $2.15 \pm 0.1$& 1.95 & 2.45 & 1.76 &
  $1.50_{-0.20}^{+0.18}$ & $2.220\pm 0.505$ \\
$\Sigma_{c}^{+}$& 0.39&0.33& $0.46 \pm 0.03$& 0.41 & 0.25 & 0.36 &
$0.12_{-0.10}^{0.06}$ & -- \\ 
$\Sigma_{c}^{0}$& -$0.79$&-$0.94$& -$1.24 \pm 0.05$& -$1.1$ & -$1.96$
& -$1.04$ & -$1.25_{-0.08}^{+0.08}$ & -$1.073\pm 0.269$ \\
$\Xi_{c}^{\prime +}$&0.39&0.44& $0.60 \pm 0.02$& 0.77 & -- & 0.47 &
$0.32_{-0.11}^{+0.13}$ & $0.315\pm0.141$ \\ 
$\Xi_{c}^{\prime 0}$&-$0.79$&-$0.84$& -$1.05 \pm 0.04$& -$1.12$
&--&-$0.95$ & -$0.95_{-0.05}^{+0.08}$ & -$0.599\pm0.071$ \\ 
$\Omega_{c}^{0}$&-$0.79$&-$0.75$& -$0.85 \pm 0.05$& -$0.79$ &--&
-$0.85$ &-$0.67_{-0.09}^{+0.09}$ & -$0.639\pm 0.088$ \\ 
 \hline 
 \hline
\end{tabular}
\label{tab:3}
\end{table}
\begin{table}[htp]
  \centering
\caption{Magnetic moments of the bottom baryon sextet with spin 1/2
in units of the nuclear magneton $\mu_N$.} 

\begin{tabular}{ c  c c  |  c  c c } 
 \hline 
 \hline 
& $\mu_{B_b}^{(m_{\mathrm{s}}=0 \,\mathrm{MeV})}$  &
$\mu_{B_b}^{(m_{\mathrm{s}}=166\,\mathrm{MeV})}$  
& \cite{Yang:2018uoj} 
& \cite{Scholl:2003ip}
& \cite{Faessler:2006ft}   \\  
 \hline
$\Sigma_{b}^{+}$ & 1.58   &1.60& $2.15 \pm 0.1$ &2.52 & 2.07\\
$\Sigma_{b}^{0}$& 0.39&0.33& $0.46 \pm 0.03$ & 0.29 & 0.53\\
$\Sigma_{b}^{-}$& -$0.79$&-$0.93$& -$1.24 \pm 0.05$ &-$1.94$ & -$1.01$\\ 
$\Xi_{b}^{\prime 0}$&0.39&0.44& $0.60 \pm 0.02$ & --& 0.66 \\
$\Xi_{b}^{\prime -}$&-$0.79$&-$0.84$& -$1.05 \pm 0.04$ &--& -$0.91$ \\
$\Omega_{b}^{-}$&-$0.79$&-$0.75$& -$0.85 \pm 0.05$ &--& -$0.82$\\
 \hline 
 \hline
\end{tabular}
\label{tab:4}
\end{table}
While there are very few theoretical works on the EM form factors of
the heavy baryons~\cite{Liu:2018tqe, Liu:2016wzh}, the magnetic
moments of the heavy baryons have been studied within various
theoretical models. In this work, we compare the present results of 
the magnetic moments with those of Refs.~\cite{Yang:2018uoj,Oh:1991ws,
Scholl:2003ip,Faessler:2006ft,Wang:2018gpl,Can:2013tna,
Bahtiyar:2016dom}. In Table~\ref{tab:3}, we list the 
results of the magnetic moments of the baryon sextet with spin 1/2.
In the second and third columns, the results of the magnetic moments
without and with the linear $m_{\mathrm{s}}$ corrections,
respectively. The effects of the $\mathrm{SU}_{\mathrm{f}}(3)$
symmetry breaking are in the range of $2\%-20\%$. For example, the
$\Omega_c^0$ magnetic moment acquires a marginal contribution from the
$m_{\mathrm{s}}$ corrections, whereas $\Sigma_c^0$ gets about $20\%$
corrections. We first compare the results with those from 
Ref.~\cite{Yang:2018uoj} where the same framework was used but the
dynamical parameters $w_i$ were fixed by using the experimental data
on those of the baryon octet. We see that the magnitudes of the
results are consistently smaller than those of
Ref.~\cite{Yang:2018uoj}. Except for the $\Omega_c^0$ magnetic moment,
the present results are in general smaller than those of
Ref.~\cite{Oh:1991ws} where the Skyrme model with the bound-state
approach was employed. Reference~\cite{Scholl:2003ip} extended the
model used in Ref.~\cite{Oh:1991ws}, including the vector mesons. The
results of the $\Sigma_c^{++}$ and $\Sigma_c^0$ turn out to be the
largest in size among all other models. On the other hand, that of
$\Sigma_c^+$ from Ref.~\cite{Scholl:2003ip} is the
smallest. Interestingly, the results of Ref.~\cite{Faessler:2006ft}
are very similar to the present ones, even though the relativistic
three-quark model of Ref.~\cite{Faessler:2006ft} is very
different from the present approach. In the final column, the lattice
results are given~\cite{Can:2013tna, Bahtiyar:2016dom}, which show
qualitatively a similar tendency. In Table~\ref{tab:4}, we list the 
results of the bottom baryon sextet with spin 1/2 for
completeness. The results are basically the same as those of the
charmed baryons. 

\begin{table}[htp]
  \centering
\caption{Magnetic radii of the charmed baryon sextet with spin 1/2 in
  units of $\mathrm{fm}^2$.} 

\begin{tabular}{  c  c  c  c} 
 \hline 
 \hline 
Baryon &  $\langle r^{2} \rangle_{M}^{B_c(m_{s} = 0 \, \mathrm{MeV})}
$  &  $\langle r^{2} \rangle_{M}^{B_c (m_{s} = 174 \, \mathrm{MeV})} $ 
& \cite{Can:2013tna} \\ 
 \hline
$\Sigma_{c}^{++}$&  0.62& 0.62 & $0.696 \pm 0.153$ \\
$\Sigma_{c}^{+}$&  0.62& 0.40 &  --  \\
$\Sigma_{c}^{0}$&  0.62& 0.78 & $0.650 \pm 0.126$  \\
$\Xi_{c}^{+}$&  0.62& 0.63 & -- \\
$\Xi_{c}^{0}$&  0.62& 0.72 & -- \\
$\Omega_{c}^{0}$&  0.62& 0.64 & $0.354 \pm 0.054$  \\
 \hline 
 \hline
\end{tabular}
\label{tab:5}
\end{table}
\begin{table}
\caption{Magnetic radii of the bottom baryon sextet with spin 1/2 in
  units of $\mathrm{fm}^2$.}
  \centering
\begin{tabular}{  c  c  c  } 
 \hline 
 \hline 
Baryon &  $\langle r^{2} \rangle_{M}^{B_b(m_{s} = 0 \, \mathrm{MeV})} 
         $  &  $\langle r^{2} \rangle_{M}^{B_b(m_{s} = 166 \, \mathrm{MeV})}  $ \\ 
 \hline
$\Sigma_{b}^{+}$&  0.62& 0.62 \\
$\Sigma_{b}^{0}$&  0.62& 0.42 \\
$\Sigma_{b}^{-}$&  0.62& 0.77 \\
$\Xi_{b}^{0}$&  0.62& 0.63 \\
$\Xi_{b}^{-}$&  0.62& 0.71 \\
$\Omega_{b}^{-}$&  0.62& 0.64 \\
 \hline 
 \hline
\end{tabular}
\label{tab:6}
\end{table}
The magnetic radius is defined by 
\begin{align}
\langle r^2 \rangle_M^{B_Q} = -\frac{6}{\mu_B} \left. \frac{d
  G_M^{B}(Q^2)}{dQ^2}   \right |_{Q^2=0} .  
\label{eq:magnetic_radii}
\end{align}
In Table~\ref{tab:5}, the numerical results of
Eq.~\eqref{eq:magnetic_radii} are listed in comparison with the
lattice data. When the linear $m_{\mathrm{s}}$ corrections are
switched off, all the results turn out to be the same. This can be
easily understood. As shown in Eq.~\eqref{eq:magnetic_radii} the
magnetic radius is normalized by the magnetic moment of the
corresponding heavy baryon. In the present mean-field formalism, we
find that the flavor part or the $D$-function part of the derivative
of the magnetic form factor is canceled by the normalization due to
the magnetic moment. Thus, all the magnetic radii of the lowest-lying
baryon sextet with spin 1/2 are expressed by the single equation
\begin{align}
\langle r^{2} \rangle_{M}^{\Sigma^{++}_{c}}=\langle r^{2}
  \rangle_{M}^{\Sigma^{+}_{c}} =\langle r^{2}
  \rangle_{M}^{\Sigma^{0}_{c}} = \langle r^{2}
  \rangle_{M}^{\Xi^{+}_{c}} = \langle r^{2} \rangle_{M}^{\Xi^{0}_{c}}
  = \langle r^{2} \rangle_{M}^{\Omega^{0}_{c}}.   
\label{eq:mag_radii_U}
\end{align}
This is the unique feature of the model and moreover
Eq.~\eqref{eq:mag_radii_U} is a special case of the $U$-spin
relation. Note that the magnetic radii of the baryon octet do not 
satisfy this relation. The present results of the magnetic radii are
compared with those from the lattice
calculation~\cite{Can:2013tna}. Except for the $\Omega_c^0$, 
we find that the results are in qualitative agreement with the lattice 
data. Table~\ref{tab:6} lists the results of the magnetic radii for
the bottom baryon sextet with spin 1/2. As mentioned several times
already, The results have no difference from those for the charmed
baryons because of the present mean-field approach. 
\section{Summary and conclusion}
In the present work, we have investigated the electromagnetic
properties of the lowest-lying singly heavy baryons with spin 1/2
within the framework of the chiral quark-soliton model. The model is a
pion mean-field approach in which a baryon is viewed as $N_c$ valence
quarks bound by the pion mean fields created self-consistently. In the
same manner, a singly heavy baryon can be regarded as $N_c-1$ valence
quarks in the presence of the pion mean fields, its mass being assumed
to be infinitely heavy. In this limit of the infinite heavy-quark
mass, the heavy quark inside a singly heavy baryon can be treated as a
mere static color source. Thus, the structure of the heavy baryon is
mainly governed by the light-quark dynamics. 
In the chiral quark-soliton model, the constraint on the quantization
rule is imposed by the number of valence quarks. In the case of light
baryons, the right hypercharge is constrained to be $Y'=N_c/3$. Since
the singly heavy quark, however, consists of the $N_c-1$ valence
quarks with the heavy quark stripped off, the quantization rule should
be modified by $Y'=(N_c-1)/3$. Then the quantization naturally yields
the baryon antitriplet and the baryon sextet with both spins 1/2 and
3/2.  

We first studied the electric properties of the baryon antitriplet
and sextet with spin 1/2. The results show that the $Q^2$ dependence
of the electric form factors for the positve-charged heavy baryons is 
very similar. Comparison of these results with that of the proton electric
form factor points to the conclusion that the heavy baryon is an
electrically compact object. The result of the $\Sigma_c^{++}$
electric form factor was compared with the lattice data. As
anticipated, the present result falls off faster than the lattice one
as $Q^2$ increases. Keeping in mind that all the lattice results of
the proton electric form factor overestimate the experimental data
when the unphysical pion mass is employed, we are able to state that
the present mean-field approach produce the consistent results of the
electric form factors of the positive-charged heavy baryons. We also
computed the electric form factors of the neutral heavy baryons. The
results of the electric charge radii were also presented. We found
that the effects of $SU_{\mathrm{f}}(3)$ symmetry breaking are
marginal. 

Since the heavy baryons in the antitriplet contain the light-quark
pair in spin zero in the present scheme, the magnetic form factors
vanish. So, we concentrated on those of the baryon sextet with spin
1/2. The magnetic densities of the positive-charge heavy baryons are
very similar to that of the proton, whereas the neutral ones take after
the neutron one. The results show that the magnetic form factors of 
the heavy baryons fall off monotonically as $Q^2$ increases. The
results of the magnetic moments were compared with those from various
works and were found to be consistent each other, though there are
differences quantitatively. The magnetic radii were also
calculated. Interestingly, when the effects of flavor SU(3) 
symmetry breaking are turned off, all the magnetic radii of the heavy
baryons turn out to be the same. This arises from the fact that the
flavor parts of the magnetic form factors are exactly canceled by the
normalizations, i.e., magnetic moments. This can be also understood as
a special case of the $U$-spin relation. 

In conclusion, the present pion mean-field approach describes the
electromagnetic properties of the lowest-lying singly heavy baryons
consistently, compared with other models and lattice QCD. Thus, a
singly heavy baryon is indeed mainly explained by the light quarks
inside it, while the heavy quark remains as a static color source. Of
course the effects of higher-order corrections in the expansion of the
heavy quark mass should be required in order to describe the
electromagnetic properties of the heavy baryons. This is a very
interesting issue for the future works. It is also of great interest 
to study the electromagnetic form factors of the sextet baryon with
spin 3/2. As in the case of the baryon decuplet, we have additionally
more form factors for the spin-3/2 heavy baryons such as the electric
quadrupole and magnetic octupole form factors. The corresponding work
is under way.

%-------------------------------------------------
\begin{acknowledgments}
%-------------------------------------------------
The authors are grateful to Gh.-S. Yang for valuable discussion. 
They also want to express their gratitude to K. U. Can and M. Oka
for providing them with the lattice data on the electromagnetic form
factors of the heavy baryons. 
The present work was supported by Basic Science
Research Program through the National Research Foundation of Korea
funded by the Ministry of Education, Science and Technology (Grant 
No. NRF-2018R1A2B2001752).
\end{acknowledgments}

\appendix

\section{Matrix elements of the SU(3) Wigner ${D}$
  function \label{app:B}} 
In the following, we list in Tables~\ref{tab:appd1}-\ref{tab:appd8} the
results of the matrix elements of the relevant collective operators
for the EM form factors of the heavy baryons. 
   
\begin{table}[htp]
  \caption{The matrix elements of the collective operators of the
    leading terms and the $1/N_c$ rotational corrections to the
    electric form factors.} 
  \label{tab:appd1}
\begin{center}
\begin{tabular}{ c } 
 \hline 
  \hline 
$\langle \Lambda_{c} |D^{(8)}_{88} | \Lambda_{c} \rangle = \langle
  \Xi_{c} |D^{(8)}_{88} | \Xi_{c} \rangle =\frac{3}{8} Y $   \\ 
$\langle \Lambda_{c} |D^{(8)}_{38} | \Lambda_{c} \rangle = \langle
  \Xi_{c} |D^{(8)}_{38} | \Xi_{c} \rangle =\frac{\sqrt{3}}{4} T_{3} $
  \\ 
$\langle \Lambda_{c} |D^{(8)}_{8i}J_{i} | \Lambda_{c} \rangle =
  \langle \Xi_{c} |D^{(8)}_{8i}J_{i} | \Xi_{c} \rangle = \langle
  \Lambda_{c} |D^{(8)}_{3i}J_{i} | \Lambda_{c} \rangle = \langle
  \Xi_{c} |D^{(8)}_{3i}J_{i} | \Xi_{c} \rangle =0$   \\ 
$\langle \Lambda_{c} |D^{(8)}_{8a}J_{a}  | \Lambda_{c} \rangle =
  \langle \Xi_{c} |D^{(8)}_{8a}J_{a}  | \Xi_{c} \rangle
  =-\frac{3\sqrt{3}}{8} Y $   \\ 
$\langle \Lambda_{c} |D^{(8)}_{3a}J_{a}  | \Lambda_{c} \rangle =
  \langle \Xi_{c} |D^{(8)}_{3a}J_{a}  | \Xi_{c} \rangle =-\frac{3}{4}
  T_{3} $   \\ 
 \hline 
$\langle \Sigma_{c} |D^{(8)}_{88} | \Sigma_{c} \rangle = \langle
  \Xi'_{c} |D^{(8)}_{88} | \Xi'_{c} \rangle =\langle \Omega_{c}
  |D^{(8)}_{88} | \Omega_{c} \rangle =\frac{3}{20} Y$   \\ 
$\langle \Sigma_{c} |D^{(8)}_{38} | \Sigma_{c} \rangle = \langle
  \Xi'_{c} |D^{(8)}_{38} | \Xi'_{c} \rangle =\langle \Omega_{c}
  |D^{(8)}_{38} | \Omega_{c} \rangle =\frac{\sqrt{3}}{10} T_{3} $   \\ 
$\langle \Sigma_{c} |D^{(8)}_{8i}J_{i}  | \Sigma_{c} \rangle = \langle
  \Xi'_{c} |D^{(8)}_{8i}J_{i}  | \Xi'_{c} \rangle =\langle \Omega_{c}
  |D^{(8)}_{8i}J_{i} | \Omega_{c} \rangle =-\frac{3\sqrt{3}}{10} Y $
  \\ 
$\langle \Sigma_{c} |D^{(8)}_{3i}J_{i}  | \Sigma_{c} \rangle = \langle
  \Xi'_{c} |D^{(8)}_{3i}J_{i}  | \Xi'_{c} \rangle =\langle \Omega_{c}
  |D^{(8)}_{3i}J_{i} | \Omega_{c} \rangle =-\frac{3}{5} T_{3} $   \\ 
$\langle \Sigma_{c} |D^{(8)}_{8a}J_{a}  | \Sigma_{c} \rangle = \langle
  \Xi'_{c} |D^{(8)}_{8a}J_{a}  | \Xi'_{c} \rangle =\langle \Omega_{c}
  |D^{(8)}_{8a}J_{a} | \Omega_{c} \rangle =-\frac{3\sqrt{3}}{20} Y $
  \\ 
$\langle \Sigma_{c} |D^{(8)}_{3a}J_{a}  | \Sigma_{c} \rangle = \langle
  \Xi'_{c} |D^{(8)}_{3a}J_{a}  | \Xi'_{c} \rangle =\langle \Omega_{c}
  |D^{(8)}_{3a}J_{a} | \Omega_{c} \rangle =-\frac{3}{10} T_{3} $   \\ 
 \hline 
 \hline
\end{tabular}
\end{center}
\end{table}

\begin{table}[htp]
  \caption{The matrix elements of the collective operators of the
    leading terms and the $1/N_c$ rotational corrections to the
    magnetic form factors.}
  \label{tab:appd2}
\begin{center}
\begin{tabular}{ c } 
 \hline 
 \hline 
$\langle \Sigma_{c} |D^{(8)}_{33} | \Sigma_{c} \rangle = \langle
  \Xi'_{c} |D^{(8)}_{33} | \Xi'_{c} \rangle =\langle \Omega_{c}
  |D^{(8)}_{33} | \Omega_{c} \rangle =-\frac{1}{5} T_{3} $   \\ 
$\langle \Sigma_{c} |D^{(8)}_{83} | \Sigma_{c} \rangle = \langle
  \Xi'_{c} |D^{(8)}_{83} | \Xi'_{c} \rangle =\langle \Omega_{c}
  |D^{(8)}_{83} | \Omega_{c} \rangle =-\frac{3}{10\sqrt{3}} Y $   \\ 
$\langle \Sigma_{c} |D^{(8)}_{38}J_{3}  | \Sigma_{c} \rangle = \langle
  \Xi'_{c} |D^{(8)}_{38}J_{3}  | \Xi'_{c} \rangle =\langle \Omega_{c}
  |D^{(8)}_{38}J_{3} | \Omega_{c} \rangle =\frac{1}{5\sqrt{3}} T_{3} $
  \\ 
$\langle \Sigma_{c} |D^{(8)}_{88}J_{3}  | \Sigma_{c} \rangle = \langle
  \Xi'_{c} |D^{(8)}_{88}J_{3}  | \Xi'_{c} \rangle =\langle \Omega_{c}
  |D^{(8)}_{88}J_{3} | \Omega_{c} \rangle =\frac{1}{10} Y $   \\ 
$\langle \Sigma_{c} |d_{ab3}D^{(8)}_{3a}J_{b}  | \Sigma_{c} \rangle =
  \langle \Xi'_{c} |d_{ab3}D^{(8)}_{3a}J_{b}  | \Xi'_{c} \rangle
  =\langle \Omega_{c} |d_{ab3}D^{(8)}_{3a}J_{b} | \Omega_{c} \rangle
  =\frac{1}{10} T_{3} $   \\ 
$\langle \Sigma_{c} |d_{ab3}D^{(8)}_{8a}J_{b}  | \Sigma_{c} \rangle =
  \langle \Xi'_{c} |d_{ab3}D^{(8)}_{8a}J_{b}  | \Xi'_{c} \rangle
  =\langle \Omega_{c} |d_{ab3}D^{(8)}_{8a}J_{b} | \Omega_{c} \rangle
  =\frac{3}{20\sqrt{3}} Y $   \\ 
 \hline 
 \hline
\end{tabular}
\end{center}
\end{table}

\begin{table}[h]
  \caption{The matrix elements of the collective operators of the
    $m_s$ corrections to the electric form factors.} 
  \label{tab:appd3}
\begin{center}
\begin{tabular}{ c | c  c    | c  c  c   } 
 \hline 
  \hline 
 $ {\cal{R}}$& \multicolumn{2}{c|}{$\bm{\overline{3}}$} &
  \multicolumn{3}{c}{$\bm{6}$}  \\  
B & $\Lambda_{c}$ & $\Xi_{c}$ & $\Sigma_{c}$  & $\Xi_{c}'$ &
 $\Omega_{c}$ \\  
 \hline
$\langle B_{{\cal{R}}} |D^{(8)}_{8i}D^{(8)}_{3i} | B_{{\cal{R}}}
  \rangle$  & 0 & $\frac{3\sqrt{3}}{20}T_{3}$ &
 $\frac{11}{60\sqrt{3}}T_{3}$ & $\frac{1}{6\sqrt{3}}T_{3}$ & 0 \\  
$\langle B_{{\cal{R}}} |D^{(8)}_{8i}D^{(8)}_{8i} | B_{{\cal{R}}}
  \rangle$  & $\frac{9}{40}$ & $\frac{9}{20}$ & $\frac{19}{60}$ &
 $\frac{2}{5}$ & $\frac{1}{2}$ \\  
$\langle B_{{\cal{R}}} |D^{(8)}_{8a}D^{(8)}_{3a} | B_{{\cal{R}}}
  \rangle$  & 0 & $-\frac{\sqrt{3}}{10}T_{{3}}$ &
 $-\frac{2}{15\sqrt{3}}T_{{3}}$ & $-\frac{1}{15\sqrt{3}}T_{{3}}$& 0
  \\   
$\langle B_{{\cal{R}}} |D^{(8)}_{8a}D^{(8)}_{8a} | B_{{\cal{R}}}
  \rangle$  & $\frac{3}{5}$ & $\frac{9}{20}$ & $\frac{8}{15}$ &
 $\frac{1}{2}$ & $\frac{2}{5}$  \\  
$\langle B_{{\cal{R}}} |D^{(8)}_{88}D^{(8)}_{38} | B_{{\cal{R}}}
  \rangle$ & 0 & $-\frac{\sqrt3}{20}T_{3}$ &
 $-\frac{1}{20\sqrt{3}}T_{3}$ & $-\frac{1}{40\sqrt{3}}T_{3}$ & 0  \\  
$\langle B_{{\cal{R}}} |D^{(8)}_{88}D^{(8)}_{88}| B_{{\cal{R}}}
  \rangle$ & $\frac{7}{40}$ & $\frac{1}{10}$ & $\frac{3}{20}$ &
 $\frac{1}{10}$ & $\frac{1}{10}$ \\  
 \hline 
 \hline
\end{tabular}
\end{center}
\end{table}

\begin{table}[htp]
  \caption{The matrix elements of the collective operators of the
    $m_s$ corrections to the magnetic form factors.} 
  \label{tab:appd4}

\begin{center}
\begin{tabular}{ c  | c  c  c   } 
 \hline 
  \hline 
$ {\cal{R}}$ & \multicolumn{3}{c}{$\bm{6}$}  \\ 
B & $\Sigma_{c}$  & $\Xi_{c}'$ & $\Omega_{c}$ \\ 
 \hline
$\langle B_{{\cal{R}}} |D^{(8)}_{88}D^{(8)}_{33} | B_{{\cal{R}}}
  \rangle$   & $-\frac{2}{45}T_{3}$ & $-\frac{1}{45}T_{3}$ & 0 \\  
$\langle B_{{\cal{R}}} |D^{(8)}_{88}D^{(8)}_{83} | B_{{\cal{R}}}
  \rangle$   & $\frac{1}{30\sqrt{3}}$ & $0$ & $-\frac{1}{10\sqrt{3}}$
  \\  
$\langle B_{{\cal{R}}} |D^{(8)}_{83}D^{(8)}_{38} | B_{{\cal{R}}}
  \rangle$  & $-\frac{2}{45}T_{3}$ & $-\frac{1}{45}T_{3}$ & 0 \\  
$\langle B_{{\cal{R}}} |D^{(8)}_{83}D^{(8)}_{88} | B_{{\cal{R}}}
  \rangle$  & $\frac{1}{30\sqrt{3}}$ & $0$ & $-\frac{1}{10\sqrt{3}}$
  \\  
$\langle B_{{\cal{R}}} |d_{ab3}D^{(8)}_{8a}D^{(8)}_{8b} |
  B_{{\cal{R}}} \rangle$ & $\frac{2}{45}$ & $-\frac{1}{30}$ &
$-\frac{1}{15}$  \\  
$\langle B_{{\cal{R}}} |d_{ab3}D^{(8)}_{3a}D^{(8)}_{8b}| B_{{\cal{R}}}
  \rangle$ & $-\frac{1}{9\sqrt{3}}T_{3}$  &
 $-\frac{7}{45\sqrt{3}}T_{3}$ & 0 \\  
 \hline 
 \hline
\end{tabular}
\end{center}
\end{table}

\begin{table}[htp]
  \caption{The relevant transition matrix elements of the collective
    operators coming from the anti-15plet component of the baryon wave 
    functions for the electric form factors.}  
  \label{tab:appd5}
\begin{center}
\begin{tabular}{ c | c  c  | c  c  c   } 
 \hline 
  \hline 
 $ {\cal{R}}$& \multicolumn{2}{c|}{$\bm{\overline{3}}$} &
 \multicolumn{3}{c}{$\bm{6}$}  \\  
$B$ & $\Lambda_{c}$ & $\Xi_{c}$ & $\Sigma_{c}$  & $\Xi_{c}'$ &
 $\Omega_{c}$ \\  
 \hline
$\langle B_ {\bm{\overline{15}}} |D^{(8)}_{88} | B_{\cal{R}} \rangle$
 & $\frac{3}{4\sqrt{5}}$ & $\frac{3}{8}\sqrt{\frac{3}{5}}$
 &  $\sqrt{\frac{1}{10}}$ &  $\frac{1}{4}\sqrt{\frac{3}{5}}$ & 0 \\  
$\langle B_ {\bm{\overline{15}}}  |D^{(8)}_{38} | B_{\cal{R}} \rangle$
 & 0 &  $-\frac{1}{4\sqrt{5}}T_{3}$ &
 $-\frac{1}{\sqrt{30}}T_{3}$ & $-\frac{\sqrt{5}}{6}T_{3}$ & 0 \\  
$\langle B_ {\bm{\overline{15}}}  |D^{(8)}_{8i}J_{i} | B_{\cal{R}}
  \rangle$  & 0 & 0 & $\sqrt{\frac{2}{15}}$ & $\frac{1}{2\sqrt{5}}$ &
 0  \\  
$\langle B_ {\bm{\overline{15}}}  |D^{(8)}_{3i}J_{i} | B_{\cal{R}}
  \rangle$ & 0 & 0 & $-\frac{1}{3}\sqrt{\frac{2}{5}}T_{3}$ &
 $-\frac{1}{3}\sqrt{\frac{5}{3}}T_{3}$ & 0  \\  
$\langle B_ {\bm{\overline{15}}}  |D^{(8)}_{8a}J_{a} | B_{\cal{R}}
  \rangle$ & $\frac{1}{4}\sqrt{\frac{3}{5}}$ &
 $\frac{3}{8}\sqrt{\frac{1}{5}}$ & $-\sqrt{\frac{1}{30}}$ &
 $-{\frac{1}{4\sqrt{5}}}$ & 0  \\   
$\langle B_ {\bm{\overline{15}}}  |D^{(8)}_{3a}J_{a} | B_{\cal{R}}
  \rangle$ & 0 & $-\frac{1}{4}\sqrt{\frac{1}{15}}T_{3}$ &
$\frac{1}{3}\sqrt{\frac{1}{10}}T_{3}$ &
  $\frac{1}{6}\sqrt{\frac{5}{3}}T_{3}$ & 0  \\   
 \hline 
 \hline
\end{tabular}
\end{center}
\end{table}

\begin{table}[htp]
\caption{The relevant transition matrix elements of the collective
    operators coming from the anti-24plet component of the baryon wave 
    functions for the electric form factors.}
  \label{tab:appd6}
\begin{center}
\begin{tabular}{ c | c  c  | c  c  c   } 
 \hline 
  \hline 
 $ {\cal{R}}$& \multicolumn{2}{c|}{$\bm{\overline{3}}$} &
 \multicolumn{3}{c}{$\bm{6}$}  \\  
$B$ & $\Lambda_{c}$ & $\Xi_{c}$ & $\Sigma_{c}$  & $\Xi_{c}'$ &
 $\Omega_{c}$ \\  
 \hline
$\langle B_ {\bm{\overline{24}}} |D^{(8)}_{88} | B_{\cal{R}} \rangle$
 & 0 & 0 & $\frac{1}{5}$ & $\frac{1}{5}\sqrt{\frac{3}{2}}$
 & $\frac{1}{5}\sqrt{\frac{3}{2}}$ \\  
$\langle B_ {\bm{\overline{24}}}  |D^{(8)}_{38} | B_{\cal{R}} \rangle$
 & 0 & 0 & $\frac{1}{10\sqrt{3}} T_{3}$ &
 $\frac{2}{15\sqrt{2}} T_{3}$ & 0 \\  
$\langle B_ {\bm{\overline{24}}}  |D^{(8)}_{8i}J_{i} | B_{\cal{R}}
  \rangle$  & 0 & 0 & $-\frac{1}{5\sqrt{3}} $ & $-\frac{1}{5\sqrt{2}}
 $ & $-\frac{1}{5\sqrt{2}} $  \\  
$\langle B_ {\bm{\overline{24}}}  |D^{(8)}_{3i}J_{i} | B_{\cal{R}}
  \rangle$ & 0 & 0 & $-\frac{1}{30}T_{3}$ &
 $-\frac{2}{15}\sqrt{\frac{1}{6}}T_{3}$ & 0  \\  
$\langle B_ {\bm{\overline{24}}}  |D^{(8)}_{8a}J_{a} | B_{\cal{R}}
  \rangle$ & 0 & 0 & $\frac{2}{5} \sqrt{\frac{1}{3}}$ &
 $\frac{\sqrt{2}}{5}$ & $\frac{\sqrt{2}}{5}$  \\  
$\langle B_ {\bm{\overline{24}}}  |D^{(8)}_{3a}J_{a} | B_{\cal{R}}
  \rangle$ & 0 & 0 & $\frac{1}{15}T_{3}$ & $\frac{2}{15}
 \sqrt{\frac{2}{3}}T_{3}$ &  $0$ \\  
 \hline 
 \hline
\end{tabular}
\end{center}
\end{table}

\begin{table}[htp]
\caption{The relevant transition matrix elements of the collective
    operators coming from the anti-15plet component of the baryon wave 
    functions for the magnetic form factors.}
  \label{tab:appd7}
\begin{center}
\begin{tabular}{ c  | c  c  c   } 
 \hline 
  \hline 
$ {\cal{R}}$ & \multicolumn{3}{c}{$\bm{6}$}  \\ 
B & $\Sigma_{c}$  & $\Xi_{c}'$ & $\Omega_{c}$ \\ 
 \hline
$\langle B_{\bm{\overline{15}}} |D^{(8)}_{33}| B_{{\cal{R}}} \rangle$
             & $-\frac{1}{9}\sqrt{\frac{2}{5}}T_{3}$ &
 $-\frac{1}{9}\sqrt{\frac{5}{3}}T_{3}$ & 0 \\  
$\langle B_{\bm{\overline{15}}} |D^{(8)}_{83}| B_{{\cal{R}}} \rangle$
             & $\frac{1}{3}\sqrt{\frac{2}{15}}$ &
 $\frac{1}{6\sqrt{5}}$ & 0 \\  
$\langle B_{\bm{\overline{15}}} |D^{(8)}_{38}J_{3} | B_{{\cal{R}}}
  \rangle$  & $-\frac{1}{3}\sqrt{\frac{2}{15}}T_{3}$ &
 $-\frac{\sqrt{5}}{9}T_{3}$ & 0 \\  
$\langle B_{\bm{\overline{15}}} |D^{(8)}_{88}J_{3} | B_{{\cal{R}}}
  \rangle$  & $\frac{1}{3}\sqrt{\frac{2}{5}}$ & $\frac{1}{2\sqrt{15}}$
                  & 0 \\  
$\langle B_{\bm{\overline{15}}} |d_{ab3}D^{(8)}_{3a}J_{b} |
  B_{{\cal{R}}} \rangle$ & $-\frac{1}{9\sqrt{10}}T_{3}$ &
 $-\frac{1}{18}\sqrt{\frac{5}{3}}T_{3}$ & 0  \\  
$\langle B_{\bm{\overline{15}}} |d_{ab3}D^{(8)}_{8a}J_{b}|
  B_{{\cal{R}}} \rangle$ &  $\frac{1}{3\sqrt{30}}$  &
 $\frac{1}{12\sqrt{5}}$ & 0 \\  
 \hline 
 \hline
\end{tabular}
\end{center}
\end{table}

\begin{table}[htp]
\caption{The relevant transition matrix elements of the collective
    operators coming from the anti-24plet component of the baryon wave  
    functions for the magnetic form factors.}
  \label{tab:appd8}
\begin{center}
\begin{tabular}{ c  | c  c  c   } 
 \hline 
  \hline 
$ {\cal{R}}$ & \multicolumn{3}{c}{$\bm{6}$}  \\ 
B & $\Sigma_{c}$  & $\Xi_{c}'$ & $\Omega_{c}$ \\ 
 \hline
$\langle B_{\bm{\overline{24}}} |D^{(8)}_{33}| B_{{\cal{R}}} \rangle$
             & $-\frac{1}{90}T_{3}$ & $-\frac{2}{45\sqrt{6}}T_{3}$ & 0
  \\  
$\langle B_{\bm{\overline{24}}} |D^{(8)}_{83} | B_{{\cal{R}}} \rangle$
             & $-\frac{1}{15\sqrt{3}}$ & $-\frac{1}{15\sqrt{2}}$ &
 $-\frac{1}{15\sqrt{2}}$ \\  
$\langle B_{\bm{\overline{24}}} |D^{(8)}_{38}J_{3} | B_{{\cal{R}}}
  \rangle$  & $\frac{1}{15\sqrt{3}}T_{3}$ &
 $\frac{2\sqrt{2}}{45}T_{3}$ & 0 \\  
$\langle B_{\bm{\overline{24}}} |D^{(8)}_{88}J_{3}  | B_{{\cal{R}}}
  \rangle$  & $\frac{2}{15}$ & $\frac{1}{5}\sqrt{\frac{2}{3}}$ &
 $\frac{1}{5}\sqrt{\frac{2}{3}}$ \\  
$\langle B_{\bm{\overline{24}}} |d_{ab3}D^{(8)}_{3a}J_{b} |
  B_{{\cal{R}}} \rangle$ & $-\frac{1}{45}T_{3}$ &
 $-\frac{2}{45}\sqrt{\frac{2}{3}}T_{3}$ & 0  \\  
$\langle B_{\bm{\overline{24}}} |d_{ab3}D^{(8)}_{8a}J_{b}|
  B_{{\cal{R}}} \rangle$ & $-\frac{2}{15\sqrt{3}}$  &
 $-\frac{\sqrt{2}}{15}$ & $-\frac{\sqrt{2}}{15}$ \\  
 \hline 
 \hline
\end{tabular}
\end{center}
\end{table}

\section{Dynamical coefficients $w_i$ for the magnetic
  moments\label{app:magmom}} 
In Eq.~\eqref{eq:magop}, the collective operator for the magnetic
moments are defined in terms of the dynamical coefficients $w_i$ that
are expressed as 
\begin{align}
w_{1} &=  \int d^{3}z \frac{M_{N}}{3}    \left(  {\cal{Q}}_{0}
        (\bm{z})  + \frac{1}{I_{1}} 
 {\cal{Q}}_{1} (\bm{z}) - 2   M_{1}  {\cal{M}}_{0} (\bm{z})  \right)
        \cr 
w_{2} &= -  \frac{1}{I_{2}} \int d^{3}z \frac{M_{N}}{3}
        {\cal{X}}_{2} (\bm{z}) \cr  
w_{3} &= -  \frac{1}{I_{1}} \int d^{3}z \frac{M_{N}}{3}
        {\cal{X}}_{1} (\bm{z}) \cr  
w_{4} &=2\sqrt{3} M_{8} \int d^{3}z \frac{M_{N}}{3}
        \left(\frac{K_{2}}{I_{2}}{\cal{X}}_{2} (\bm{z}) 
     -{\cal{M}}_{2} (\bm{z})\right) \cr
w_{5} &= \frac{1}{\sqrt{3}} M_{8} \int d^{3}z \frac{M_{N}}{3}
        \left(\frac{K_{1}}{I_{1}}{\cal{X}}_{1} (\bm{z}) -
        {\cal{M}}_{1} (\bm{z})- {\cal{M}}_{0} (\bm{z})\right) \cr 
w_{6} &= -\frac{1}{\sqrt{3}} M_{8}  \int d^{3}z \frac{M_{N}}{3}
        \left(\frac{K_{1}}{I_{1}}{\cal{X}}_{1} (\bm{z}) -
        {\cal{M}}_{1} (\bm{z})+ {\cal{M}}_{0} (\bm{z})\right) . 
\end{align}

The results of $w_i$ are listed in Table~\ref{tab:parameter} in
comparison with those from Ref.~\cite{Yang:2018uoj}, where $w_i$ were
determined by using the experimental data on the magnetic moments of
the baryon octet. 
\begin{table}[htp]
  \centering
\caption{Magnetic $\chi$ QSM parameters for $M = 420$MeV heavy baryon.} 

 \begin{tabular}{ c | c c c c c c} 
  \hline 
    \hline 
&${w}_1$ & ${w}_2$  &  ${w}_3$ & ${w}_4$ & ${w}_5$  &  ${w}_6$ \\
  \hline 
Charmed baryon($m_{s} = 174$MeV)&$-8.49$ & $4.53$  &  $2.93$ & $-1.21$
                                         & $-0.37$  &  $0.26$ \\  
Bottom baryon($m_{s} = 166$MeV)&$-8.49$ & $4.53$  &  $2.93$ & $-1.16$
                                         & $-0.35$  &  $0.25$ \\  
Yang {\it{et al.}}~\cite{Yang:2018uoj}  &$-10.08\pm0.24$ & $4.15\pm0.93$  &
  $8.54\pm0.86$ & $-2.53\pm0.14$ & $-3.29\pm0.57$  &
 $-1.34\pm0.56$    \\  
 \hline 
 \hline
\end{tabular}
\label{tab:parameter}
\end{table}

%=========================================================

\end{document}